\pdfoutput=1
\documentclass[12pt]{iopart}

\usepackage{cite}
\usepackage{graphicx}
\usepackage{longtable}
\usepackage{multirow}
\begin{document}

\title[Review of Graphene-based Thermal Polymer Nanocomposites]{Review of Graphene-based Thermal Polymer Nanocomposites: Current State of the Art and Future Prospects}

\author{Jacob S. Lewis, Timothy Perrier, Zahra Barani, Fariborz Kargar and Alexander A. Balandin}

\address{Phonon Optimized Engineered Materials (POEM) Center, University of California, Riverside, CA 92521, USA}
\address{Materials Science and Engineering Program, Bourns College of Engineering, University of California, Riverside, CA 92521, US}
\address{Department of Electrical and Computer Engineering, Bourns College of Engineering, University of California, Riverside, CA 92521, USA}
\eads{\mailto{jlewi014@ucr.edu}, \mailto{balandin@ece.ucr.edu}}
\vspace{10pt}
\begin{indented}
\item[]\today
\end{indented}

\begin{abstract}
We review the current state-of-the-art of graphene-enhanced thermal interface materials for the management of heat the next generation of electronics. Increased integration densities, speed, and power of electronic and optoelectronic devices require thermal interface materials with substantially higher thermal conductivity, improved reliability, and lower cost. Graphene has emerged as a promising filler material that can meet the demands of future high-speed and high-powered electronics. This review describes the use of graphene as a filler in curing and non-curing polymer matrices. Special attention is given to strategies for achieving the thermal percolation threshold with its corresponding characteristic increase in the overall thermal conductivity. Many applications require high thermal conductivity of the composites while simultaneously preserving electrical insulation. A hybrid filler – graphene and boron nitride – approach is presented as possible technology for independent control of electrical and thermal conduction. Reliability and lifespan performance of thermal interface materials is an important consideration towards the determination of appropriate practical applications. The present review addresses these issues in detail, demonstrating the promise of the graphene-enhanced thermal interface materials as compared to alternative technologies. 
\end{abstract}

%
\vspace{2pc}
\\

\section{Introduction}

The extraordinary increase in transistor density in semiconductor products has revolutionized our society and introduced new challenges towards its continued progress \cite{Thompson2006}. Though the decreasing feature sizes that enables ever-increasing densification has typically brought with it per-transistor energy efficiency enhancements, this does not make up for the overall waste heat production resultant from having more switches in total in the same area \cite{Krishnan2007, Lin2008}. This has led to a general trend for very large scale integration (VLSI) chips to increase in thermal design power (TDP) at every generation, with notable deviations from this trend usually coming in the form of vast architectural improvements or splitting the die into multiple logical cores. The increase in dissipated heat is problematic for VLSI semiconductor chips because their functionality can unacceptably alter at high temperatures, due for instance to hot carrier degradation and bias temperature instability \cite{Yu2018, Guo2017, Bury2018}. Now that devices are manufactured in the sub-10-nanometer process, it is becoming more difficult to manage waste heat production due to ever more important factors like leakage current and Joule heating in interconnect circuit elements of decreasing cross-sectional area. Each of these serve to make improved thermal dissipative solutions increasingly essential. In parallel, the growing fields of LED lighting and solar energy along with continuation of aerospace products all require similar and improved heat dissipation solutions \cite{Arik2004, Tang2016, Cho2016, Zhang2016b, Saadah2017, Barako2018, Mahadevan2019, Song2020}. 

The scale of the waste heat problem in semiconductors is often lost in the numbers even among researchers. The \it average \rm power density of some modern silicon VLSI chips can reach as high as 1/100 of the power density at the top of the sun's photosphere, which is approximately 6,300 W/cm$^{2}$. However, when one takes a more detailed look at a modern VLSI chip they will find local spots in which the heat density is substantially higher than the average \cite{Lorenzini2019}. VLSI chips operate at such reasonable temperatures despite their staggering heat production solely because of their accompanying engineered thermal dissipation solutions.

The most common technique to remove heat from VLSI chips and other semiconductor circuits is to bring metals -- termed heat sinks -- in contact with the chip so the heat may diffuse into this additional component. Then the heat sink would be cooled down by the environment with a presumably infinite thermal reservoir capacity. Often the heat sink employs heat pipes -- sealed tubes with an often phase-changing fluid inside of it -- that add convection and heat of vaporization at each end as mechanisms of heat transfer along with the conduction of the metallic pipe material \cite{Kang2006, Boukhanouf2006}. The heat sink class of thermal dissipation solutions are cheap, reliable, small, and ubiquitous. 

All thermal dissipation solutions in which a solid heat-producing device is placed in contact with a solid heat sink suffer from a physical junction thermal interface resistance. Between any two solid, non-compliant materials the total percentage of surface area making contact can be quite low, with a strong dependence on factors such as microscopic scale surface roughness, material plasticity, and mounting pressure \cite{Greenwood1966,Jones1968, Cooper1969}. A low proportion of direct surface contact at a physical junction inevitably means that gaps are filled with air, which has very poor heat transfer characteristics relative to the metals on each side of the junction. The heat flow from source to drain is analogous to and often thought of as an electrical circuit, in which the metal components of the dissipative solution are low resistance wires with the junction thought of as a resistor. The thermal resistance of between two physical junctions is often termed as contact resistance, $R_{C}$. In a junction with a TIM the resistance is then:

\begin{equation}
R_{TIM} = \frac{BLT}{K} + R_{C1} + R_{C2}
\end{equation}

where $BLT$ is the bondline thickness, $K$ is the thermal conductivity (TC) of the TIM itself, and $R_{C1}$ and $R_{C2}$ are the contact resistances of each junction surface with the TIM \cite{Prasher2001, Sarvar2006, Razeeb2018}. For an appropriate TIM, $R_{TIM} < R_{C}$. It is clear from Equation 1 that for increasing BLT the TC becomes an ever more important factor in $R_{TIM}$. 

The resistance of the junction is typically reduced with the use of an interstitial material called a thermal interface material (TIM) to take the place of air \cite{Burger2016}. The thermal resistance of the junction can be substantially reduced in this manner but thus far has not been comparable to the ideal of uninterrupted copper with no junction. Figure (1a) shows a schematic highlighting the benefit of TIMs in an exaggeratedly imperfect junction in which a greater portion of the junction's surface area is used for heat dissipation with TIMs applied versus without. 

Metal TIMs to date have achieved the lowest thermal interface resistance. There can be variations in precisely the functionality of this class of TIM, but they typically are introduced to the junction as a hot liquid and are frozen to a solid between the two surfaces. However, they can exist in either a permanently liquid state or alter between the two. Metal TIMs can achieve a TC over 86 W/mK -- that of Indium -- and an interfacial resistance of 0.005 Kcm$^{2}$/W \cite{Deppisch2006, Gao2012, Roy2015, Nagabandi2018}. The thermal transport in metallic TIMs is predominantly contributed to by their substantial population of free electrons, as in all metals, carrying heat mostly freely within the material's spatial confines. Though these TIMs remain at the time of this article as the best-performing at application, they are marred with reliability problems and are more expensive than alternatives. Due to the reliability concerns of metallic TIMs, it is a very active area of research for the materials \cite{Macris2004, Subramanian2005, Deppisch2006, Otiaba2014, Too2009, Ekpu2014, Roy2016}. The reason for this poor lifespan performance of metal TIMs is that they freeze into a solid that has a different coefficient of thermal expansion (CTE) with the materials on either side of it. As the temperatures of the junction are inevitably varied a disadvantageous thermal stress is inadvertently applied to the TIM and eventually cracks it, leading to substantially reduced performance. That same thermal expansion mechanic can result in pushing fluid TIMs out of the junction in a process called ``pumping out." This can be very problematic in the more modern, permanently fluid metal TIMs because there is a risk of spilling onto electrical components susceptible to electrical shorting failures. Another common class of TIMs are the elastomeric thermal pads. These TIMs are a very spongy and flexible solid pad that pushes itself into gaps in the junction due to its resistance to mechanical deformation. The highest TC achieved in this class of TIM in industry known by the authors is 62.5 W/mK. Though the TC of these are impressive, they suffer from large contact resistances that ultimately leads to a modest overall thermal resistance. 

\begin{figure}
	\centering
	\includegraphics[scale = .85]{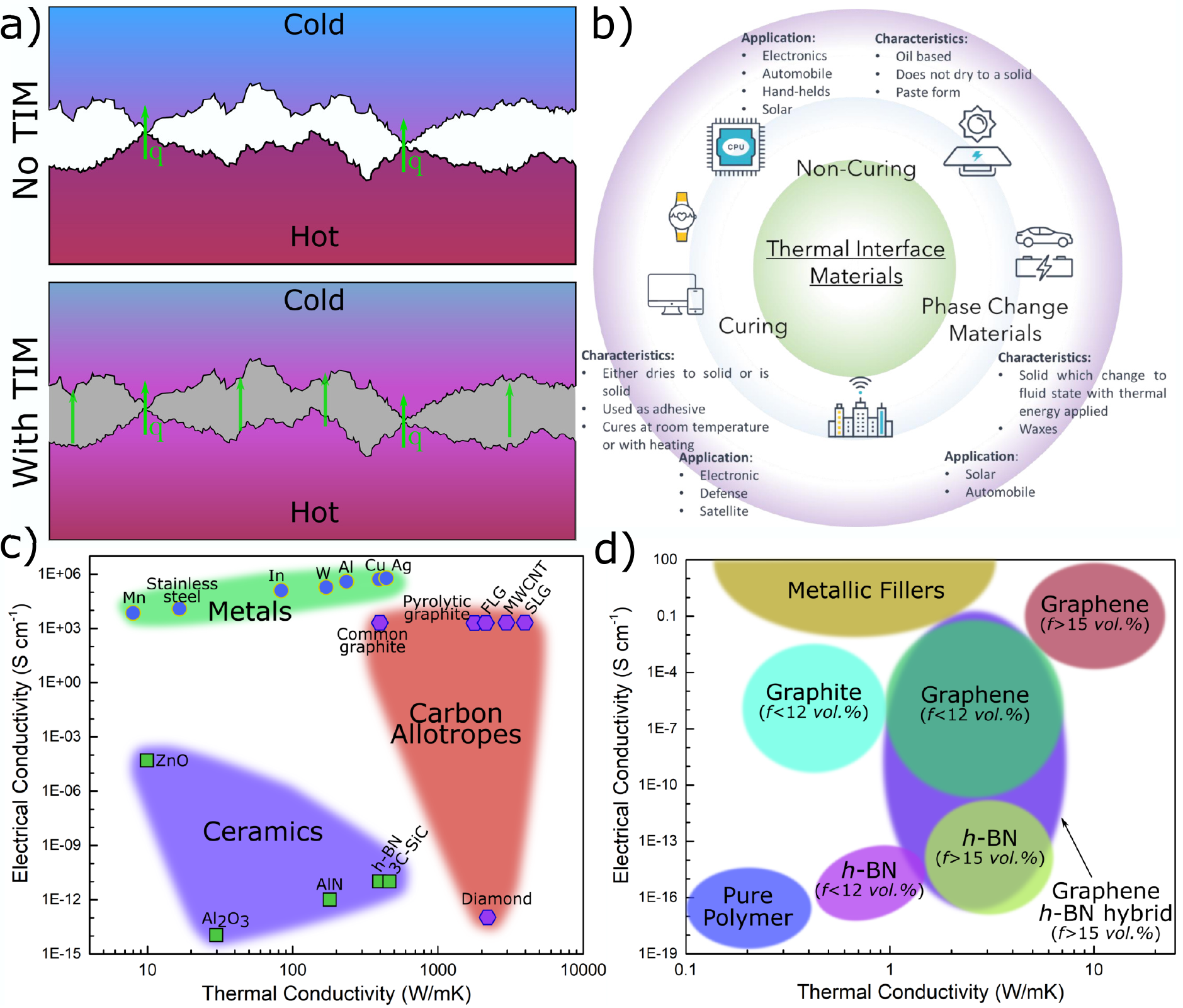}
	\caption{a) Top: a primarily air-gapped exaggerated physical interface in which noteworthy heat dissipation only occurs at a small point of contact. Bottom: The same junction after a TIM has been applied allowing substantially more heat dissipation over the otherwise air-gapped regions. b) Uses for different types of TIMs. c) Material properties of a selection of popular filler materials. d) Typical composite properties for different TIMs with un-oriented fillers. Panel \textbf{b} adapted with permission from ref \protect{\cite{Naghibi2020b}}. Published under CC License by UC Riverside.}
\end{figure}

Similarly, a solid polymer or clay material can be used in the direct encapsulation of less complex semiconductor circuits than modern VLSI chips for protection from environmental contaminants. Though chips encapsulated in this manner will have less heat-producing circuit elements than in VLSI chips, devices of this class can include high-power GaN amplifiers with substantial lifespan sensitivity to operating temperature \cite{Trew2009}. The thermal performance of the encapsulation material is an important parameter, analogous to a classic TIM. Because of this, chip encapsulation materials are considered a type of TIM. Encapsulation TIMs are typically even more sensitive to electrical conductivity (EC) due to their direct contact with active circuit elements \cite{Huang2011}. Figure (1b) shows different types of TIMs and the applications in which they are typically used.

TIM materials are often composed of metal solders, mechanically compliant pads, and polymers typically composited with filler materials \cite{Chung2001}. Each type of TIM has its own strengths and weaknesses. By far the most common class of TIM is that of the polymeric type. These TIMs have a polymer matrix in which a highly thermally conductive filler is almost always added to form a composite. This class of TIMs have a higher thermal resistance than metal-based TIMs but benefit from being stable at higher temperatures and substantially simpler to work with, especially when re-application is necessary. To date, these TIMs tend to have a lower TC than thermal pads, with a bulk TC in industry between 0.5 W/mK and 7.0 W/mK at high filler concentration, but have much less contact resistances, leading to overall slightly better performance \cite{Gwinn2003}. It should be noted that the BLT and contact resistances are influenced by the TIM's rheological properties, particularly viscosity, and often increasing the filler loading, thus $K$, of the composite comes at the sacrifice of larger BLT, $R_{C1}$, and $R_{C2}$.

Polymeric TIMs have seen considerable research into potential materials that could be used as conductive fillers. Some common polymers used are mineral and silicone oil, epoxy, poly(methyl methacrylate) (PMMA), and polyethylene \cite{Yunsheng1999, Csaba2007, Zhou2011}. Performance of base polymers can vary widely by preparation. For instance, varying the stoichiometric ratio of diglycidyl ether of bisphenol-A (DGEBA) -- a common type of epoxy used in this field of research -- can result in a factor of two alteration in its thermal diffusivity \cite{Almeida1998}. One constant requirement of all filler materials is that their physical dimensions must be small enough that a consistent mixture may be formed within the TIM. Filler materials either in industrial or research use include silver, copper, Al$_{2}$O$_{3}$, AlN, boron nitride, ZnO, diamond, graphite, carbon nanotubes, few-layer graphene (FLG), and many others \cite{Wong1999, Murshed2005, Sim2005, Cola2007, Zeng2009, Yu2011, Burger2015, Yu2015d, Du2017, Quinton2018, Devananda2019, Theerthagiri2019, Sharma2020}. A selection of works into these filler materials is summarized in Table 1 at the end of this article. Figure (1c) shows the bulk material properties for a selection of potential filler materials. For each specific geometry of filler, there exists a maximum practical filler loading that can be achieved often called the workability limit due to an unacceptable increase in composite viscosity \cite{Levy2019, OhayonLavi2020}. High TIM viscosity can complicate preparation and result in ever-increasing contact resistance in a junction. The resulting composite thermal and electrical properties that is typical for composites with randomly oriented fillers of a particular species is shown in Figure (1d).

Research into graphene-filled polymeric TIMs have flourished after the discovery of graphene's extraordinary thermal conductivity ranging from 2000 to 5300 W/mK \cite{Balandin2008, Balandin2011, Ghosh2008, Seol2010, Cai2010, Wang2016d, Zhang2017, Li2017b, Balandin2020}. Early studies showed graphene-filled TIMs with thermal conductivities as high as 5 W/mK at room temperature (RT) with graphene filler loading fractions of around 10 \it vol. \%\rm, further spurring graphene TIM research \cite{Shahil2012, Fu2014}. More recent studies into randomly-oriented graphene TIMs in a cured epoxy polymer matrix have achieved thermal conductivities of $\approx$12 W/mK \cite{Kargar2018, Shtein2015a, Kargar2019}. Counter-intuitively but interestingly, graphene has been included into aerogel and displayed a sharp and unprecedented \it reduction\rm in TC to between 4.7 $\times 10^{-3}$ and 5.9 $\times 10^{-3}$ W/mK at RT, though these results are far from typical for graphene composites \cite{Xie2016}. Graphene has promising potential in developing the next generation of TIMs. In a closely-related vein of research, graphene has been composited with thermosetting plastics with the intention to increase the polymer's fracture resistance, often with little consideration for the composite thermal properties \cite{Atif2016}.

From a practical standpoint, graphene has the potential to be a cheap filler material due to its composition of abundant carbon, given maturity in synthesis techniques. Liquid-phase exfoliation has stood out as a promising graphene synthesis method with the potential for future economic scaling \cite{Hernandez2008, Lotya2009, Coleman2011, Nicolosi2013}. This technique employs a high energy sonicator to vibrate the layers of a thick stack of graphite bound by weak van der Waals forces suspended in a fluid apart into few-layer graphene. Another interesting and scalable technique is electrochemical exfoliation in which bulk graphite is used as an electrode and solute ions intercalate into the graphite. This intercalation results in inter-layer stretching that either leads directly to exfoliation or leads to easier exfoliation when a sonication is applied \cite{Yu2015}. This technique also affords the ability for easy functionalization of the resulting graphene flakes. It is also very common and economical to oxidize graphite into graphite oxide via Hummers' Method, liquid-phase exfoliate the much simpler oxide, then finally reduce the resulting graphene oxide to a form of pure graphene \cite{Hummers1958, Stankovich2006, Stankovich2007, Nethravathi2008, Chen2013, Zaaba2017}. However, these processes have drawbacks primarily resulting in defects degrading advantageous properties of the graphene, with substantial defects in graphene derived from the reduction of graphene oxide \cite{Nagyte2020, Eigler2012, Chen2009, Hao2011, Mortazavi2013, Malekpour2016, Zhang2012}. 

In TIM research, the term ``graphene" refers to a mix of single-layer graphene (SLG) and FLG up to a few nanometers in thickness \cite{Paton2014}. Graphene's in-plane thermal conductivity is reduced with increasing layers up until $\approx$8 total monolayers, at which point the TC stabilizes to that of high-quality graphite at $\approx$2000 W/mK but still remains more mechanically flexible \cite{Nika2009, Nika2009b, Nika2012}. However, the thermal conductivity reduction resultant from contact between graphene and a dissimilar material is far more dramatic. Though there is a reduction of intrinsic TC for increasing graphene layers, there is a competing mechanism to consider where in FLG the outer layers of graphene can insulate interior layers from the substantial TC degradation from phonon scattering resultant from contact to other materials, in this case polymer matrix \cite{Jang2010, Pettes2011, Sadeghi2013, Renteria2014, Sun2016, Shen2016b, Correa2017}. The two-dimensional geometry of the graphene is an important factor leading to composites composed of graphene having typically much better TC enhancement relative to the one-dimensional carbon nanotube. However, it is important that the graphene exist in the composite with little bending lest it suffer a substantial reduction in performance \cite{Chu2013, Li2014}.

Many applications require TIMs with electrically insulating properties. Polymer TIMs can vary widely in their EC depending primarily on the type, concentration, and morphology of the filler used. An electrically conductive filler material can be used to fill a polymer TIM for such an application up to a certain level -- termed the electrical percolation threshold -- where the overall EC of the composite raises orders of magnitude, as seen in Figure (2a) \cite{Schueler1997, Sandler2003, Gelves2005, Bauhofer2009, White2010}. 

\begin{figure}[h]
	\centering
	\includegraphics[scale=0.6]{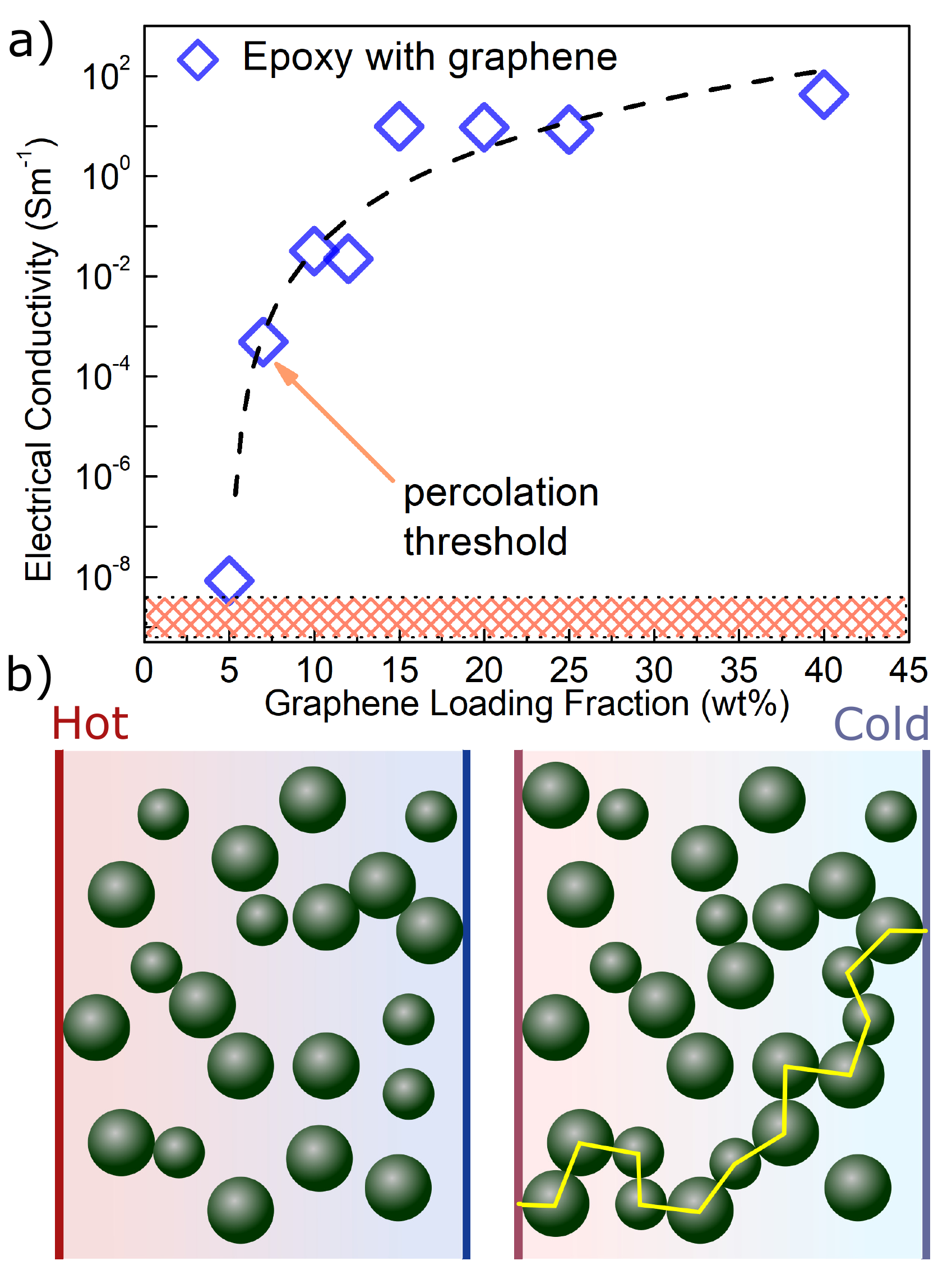}
	\caption{a) EC of a composite above and below the electrical percolation threshold with electrically conductive graphene fillers. b) Left: A TIM between a hot and cold surface with low filler loading with natural size variations. Right: The same scenario with more fillers and the development of a low-resistance percolation pathway. Panel \textbf{a} adapted with permission from ref \protect{\cite{Barani2020}}. Copyright 2019 John Wiley \& Sons.}
\end{figure}

Of considerable importance to TIMs, electrical percolation threshold has analogous behavior in TC known suitably as thermal percolation threshold. The percolations of these two material parameters are governed by the concentration and morphology of filler material required for large-scale, uninterrupted paths to become opened up from one filler particle to the next. At this point, a low resistance pathway, be it thermal or electrical, from one end of the TIM to the other becomes available and each respective property enhances substantially. Figure 2 shows two idealized hot and cold surfaces with a filler material between them. In the left schematic, the concentration of spherical fillers is low enough that most fillers are isolated from one another. In the schematic on the right the concentration is high enough that fillers make contact, making long-range contact with one another allowing for a low-resistance pathway between the two surfaces. A common trend in research is to add a filler material with poor EC to allow the use of superior thermally conductive but also electrically conductive filler without an unacceptable increase in overall TIM EC \cite{Cui2011}.

This paper covers recent advances in the promising graphene and graphene/boron nitride hybrid filled TIMs. A greater depth discussion of the thermal percolation threshold and role that adding different types of fillers -- often known as hybrid, binary, tertiary, etc. filling -- can have on it. Also considered is the all-too-often overlooked lifespan performance of these TIMs. 

\section{Recent advances of graphene TIMs}

Some of the most thermally conductive polymeric TIMs have employed the quasi-2D graphene as filler material, occasionally including a second filler as an additional component. Normally, filler materials are in general randomly-oriented by a classic mixing procedure. This random orientation of fillers is less efficient than if directionally-selective processes were employed considering that the latter scenario serves to effectively increase the size of the flake and thus unobstructed pathway along a desired direction in dimensionally constrained fillers. Studies concerned with selectively aligning graphene fillers have proven to be useful in increasing TC improvement per graphene loading level efficiency \cite{Liang2011}.

The first work on TIMs with graphene-like materials used as a filler known to the authors was conducted in 2006 \cite{Fukushima2006, Stankovich2006, Yu2007}. This work started with typical, macroscopic graphite that was oxidized and then exfoliated. The thickness of the obtained filler material was $\approx$10 nm with lateral dimensions of $\approx$15$\mu$m, a geometric portfolio typically referred to as ``few-layer graphene" today.

\begin{figure}[h]
	\centering
	\includegraphics[scale=0.4]{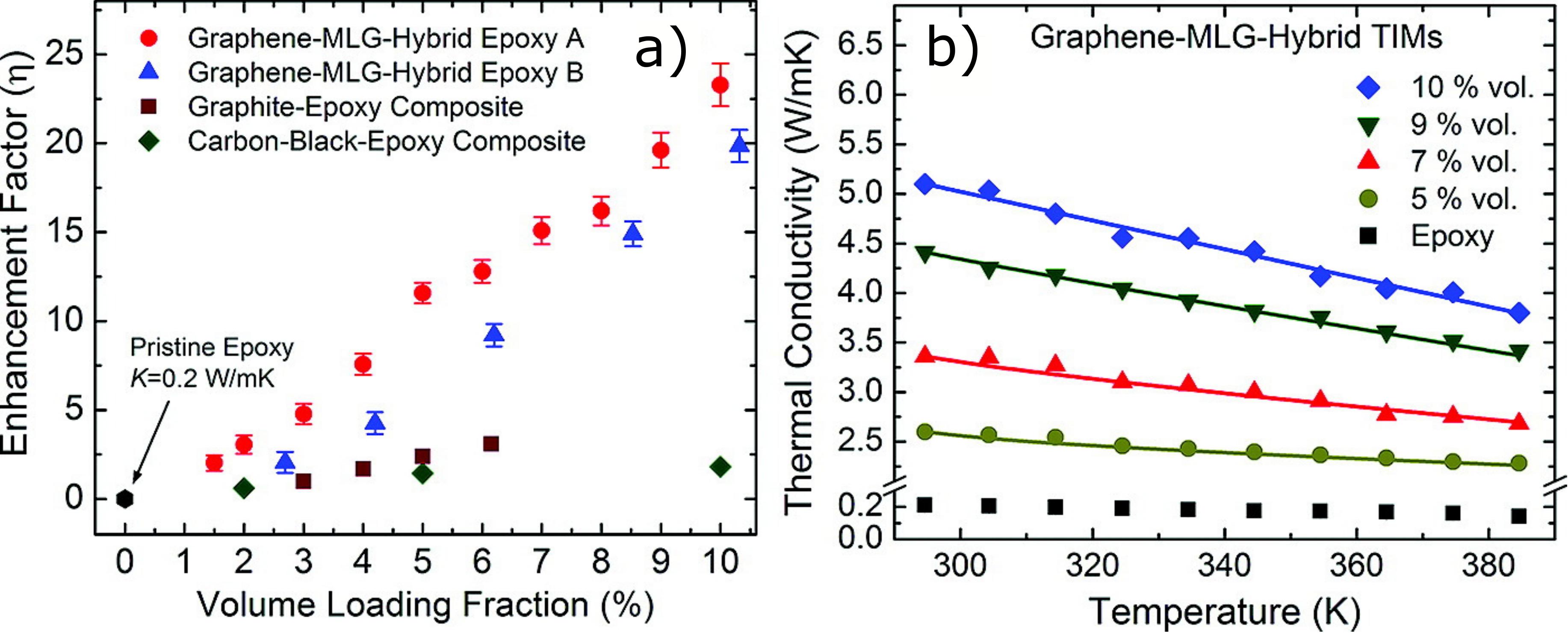}
	\caption{a) Enhancement of pure epoxy with increasing load level up to 10 $vol. \%$. ``Graphene-MLG-Hybrid Epoxy A" corresponds to a composite that was mixed for $\approx$12 h at 15,000 RPMs and ``Graphene-MLG-Hybrid Epoxy B" corresponds to a less mixed composite that went through $\approx$10 h of 5,000 RPMs of mixing. b) Temperature-dependent TC for graphene and few-layer graphene TIMs at different load levels. Adapted with permission from ref \protect{\cite{Shahil2012}}. Copyright 2012 American Chemical Society.}
\end{figure}

Tremendous interest in graphene as a filler of TIMs followed an early demonstration of a TC enhancement of 2300 \% at only 10 \it vol. \%\rm filler loading in an epoxy matrix \cite{Shahil2012}, shown in Figure (3). These results have since been confirmed by independent studies \cite{Fu2014, Shtein2015a}. Also studied was an unprecedented enhancement of a commercial TIM from $\approx$5.8 W/mK to 14 W/mK with a small addition of 2 \it vol. \%\rm of graphene. The Maxwell-Garnett effective medium approximation that is known to be effective for lower loading fractions was used to analyze the data \cite{Nan1997, Xie2008}. By treating graphene and carbon nanotubes as dramatically oblate and prolate spheroids, respectively, superior TC of graphene composites is effectively modeled. Following is the derived expression for a graphene-filled composite's TC

\begin{equation}
K = K_{p} \Bigg[\frac{3 K_{m} + 2 f (K_{p} - K_{m})}{(3 - f) K_{p} + K_{m} f + \frac{R_{B} K{m} K_{p} f}{H}}\Bigg]
\end{equation}

where $R_{B}$ is the microscopic interfacial resistance between graphene and matrix, $K_{p}$ is the TC of the flakes, $K_{m}$ is the TC of the matrix, $f$ is the loading fraction, and $H$ is the thickness of the flakes.

In all TIMs one must consider the microscopic interfacial (Kapitza) resistance of fillers within the material, a situation quite analogous to the macroscopic contact resistance that the TIM is employed to ameliorate. There is an unfortunate mismatch of phonon vibrational frequencies between graphene and polymer matrix that functionalization can address \cite{Lin2013, Vasiraju2017}. Research has been conducted to decrease the microscopic filler interfacial resistance in graphene TIMs through a functionalization process of the fillers \cite{Rohini2015, Speranza2019}. Using this technique, a TC of 1.53 W/mK in an epoxy resin polymer was achieved with 10 \it wt. \%\rm of functionalized graphene \cite{Song2013}. It was shown in molecular dynamics simulations, effective medium theory, and others that the reduction of microscopic filler interfacial resistance resulted in an increase of overall composite TC \cite{Wang2014, Wang2015b, Wang2016a, Liu2014, Zabihi2016}. Figure (4a) and (4b) show schematics of a linear hydrocarbon chain grafted to a graphene sheet to produce a functionalized surface. In Figure (4c) the thermal conductivity, $K^{*}$, of a simulated composite is analyzed at varied graphene lateral dimensions with different hydrocarbon areal densities, $\sigma$, on the graphene flakes. Interestingly, the functionalized graphene composites achieved higher TC until a filler length of $\approx$5 $\mu$m, at which point the non-functionalized graphene composite began to perform better. Alternatively, graphene functionalization can be useful to prevent agglomerations and to attach components that can be used to orient the graphene flake \cite{Renteria2015}.

\begin{figure}[h]
	\centering
	\includegraphics[scale=0.4]{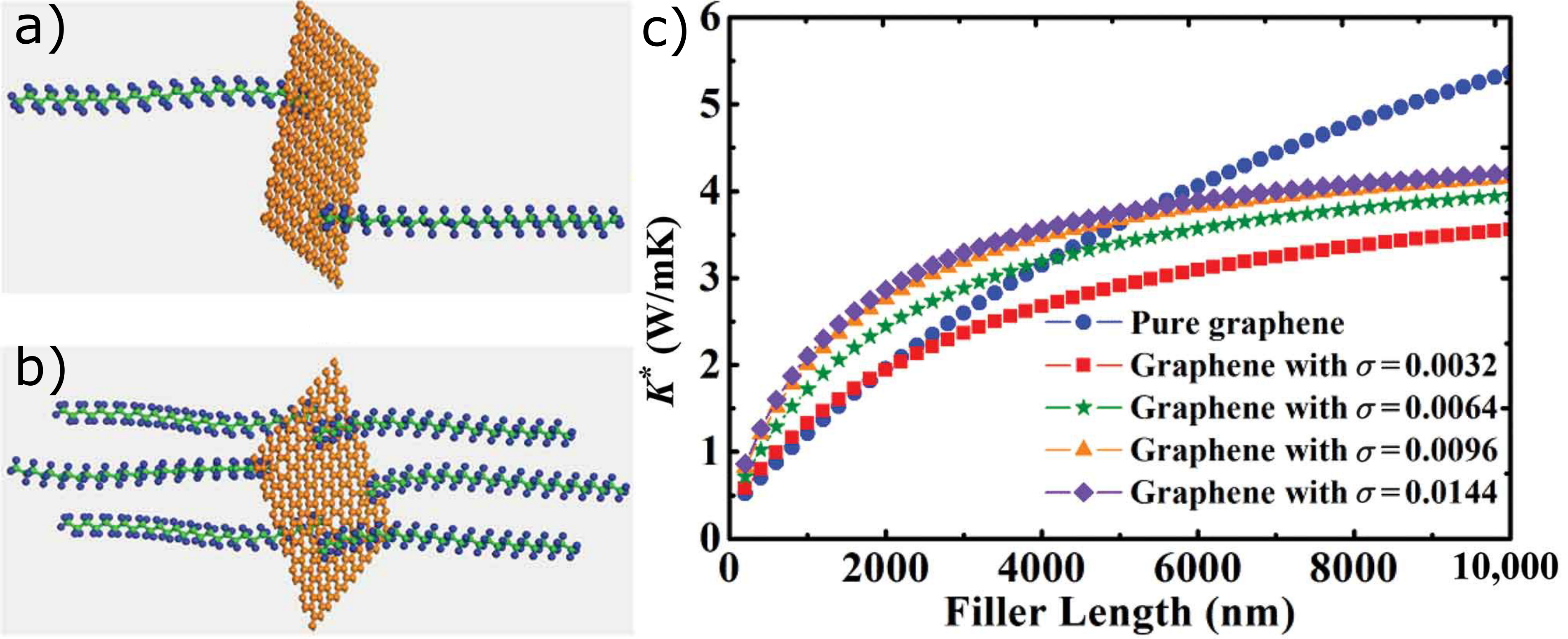}
	\caption{a) Section of a graphene flake with 2 linear hydrocarbon chains grafted on. b) 6 linear hydrocarbons grafted to graphene. c) Non-equilibrium molecular dynamics simulation composite TC with different areal densities of grafted hydrocarbons ($\sigma$) per square Angstrom. Increasing density of hydrocarbon attachments results in better performance until the length of graphene approaches approximately 5 microns, at which point the intrinsic graphene TC reduction becomes a more dominant mechanism in the composite. Adapted with permission from ref \protect{\cite{Wang2014}}. Published under the CC license by Taylor \& Francis.}
\end{figure}

Using typically very defected graphene derived from the reduction of graphene oxide, an improvement of 0.196 W/mK to 0.416 W/mK was seen in a polyamide with a graphene loading of 10 \it wt. \%\rm \cite{Ding2014}. In this study, a surface functionalization process was conducted that helped to increase the thermal coupling between the reduced graphene oxide and the polymer matrix. Using a similar reduced graphene oxide at only 1.5 \it wt. \%\rm and an additional functionalization step, a silicone matrix composite achieved a TC of 2.7 W/mK \cite{Zhang2016b}. This TIM was then applied to bridge an LED chip and a heat sink with a smaller temperature difference between the two when the TIM TC increases. In a very similar study, a graphene derived from graphene oxide and polyamide composite achieved 5.1 W/mK with functionalization and 3.34 W/mK without functionalization at 5 \it wt. \%\rm \cite{Cho2016}. Functionalization has been applied to graphene composites using gallic acid to attach a monomer and help with the dispersion of graphene in DGEBA \cite{Cao2013}. In a similar research strategy, functionalization has been used to attach silver particles to graphene to also prevent graphene from agglomerating in the composite \cite{Chen2016}. The contribution of the functionalization process to TC enhancement can be seen most starkly when comparing the prior results to one in which an epoxy polymer was filled with 2 \it wt. \%\rm non-functionalized reduced graphene oxide and achieved a very modest enhancement from 0.18 W/mK to 0.24 W/mK \cite{Olowojoba2016}. 

Researchers have used graphene functionalization to attach magnetic particles, such as Fe$_{3}$O$_{4}$, to the sheets. Then, once the functionalized graphene is dispersed within the polymer a magnetic field is applied. Because the graphene sheets are attached to them, they are aligned along the magnetic field, leading to the ability to increase the thermal transport along a particular direction. In a study with an epoxy polymer matrix, the addition of 1 \it vol. \%\rm randomly oriented graphene raised the TC of the composite from 0.17 W/mK to 0.41 W/mK \cite{Yan2014}. However, when the graphene was functionalized with Fe$_{3}$O$_{4}$ and magnetically aligned the composite achieved a rough TC of 0.57 W/mK when aligned parallel to the direction of thermal characterization and 0.25 W/mK when perpendicular. These results were verified later and shown in Figure (5f) that orienting graphene in this manner is more efficient at enhancing the TC than when using a random orientation approach \cite{Renteria2015}. Alternatively, alignment of graphene has been achieved by a clever use of interfaces between two different polymer materials to preferentially trap graphene sheets at the interface \cite{Huang2016}. This serves to both locally increase the loading level and allow for directional orientation along the interface. Another intentional filler orientation work reported a TC of 2.13 W/mK, an enhancement of 1231\%, with only 0.92 \it vol. \%\rm of graphene \cite{Lian2016}.  Recently, graphene alignment by way of a freeze-casting method that uses ice crystals to preferentially orient the flakes has grown in popularity \cite{An2018, Li2018, Liang2019, Bo2019, Wang2020}. An interesting technique to realize semi-controllable graphene orientation is to fix graphene to a 3-D structure, with a morphology similar to sponges, then cure the graphene with or without the scaffold in a polymer of choice \cite{Liu2019a, Song2019a}.

\begin{figure}
	\centering
	\includegraphics[scale=0.8]{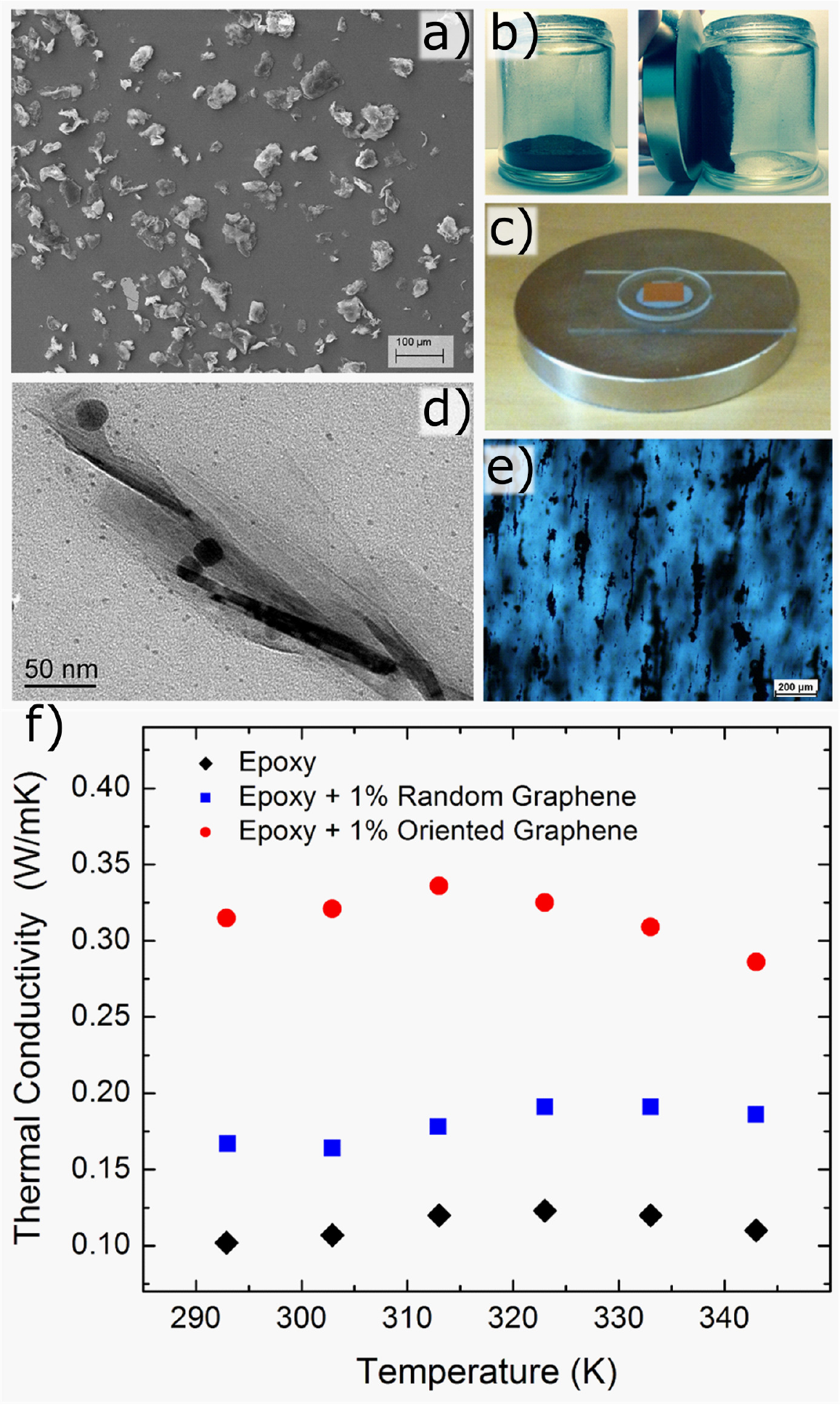}
	\caption{a) SEM image of graphene derived by liquid-phase exfoliation. b) Photograph showing the response to the magnetically-functionalized graphene powder to an applied magnetic field. c) Functionalized graphene between two copper foils and placed on a permanent magnet for filler alignment. d) TEM micrograph of graphene flake with attached Fe$_{3}$O$_{4}$. e) Optical microscopy image with low concentration of aligned filler. f) Apparent thermal conductivity at different temperatures. The superiority of intentionally oriented graphene flakes to randomly orientated graphene and pure epoxy is evident. Adapted with permission from \protect{\cite{Renteria2015}}. Copyright 2015 Elsevier.}
\end{figure}

The directional control of graphene fillers is primarily of interest because its potential to achieve order-of-magnitude improvement over current composites in the cross-plane direction (from source to sink). Selective alignment along the plane of a TIM remains an area of important inquiry but has less immediately practical implications as these composites are not well-suited for passing heat along a thin interface \cite{Kumar2016}. General TIM composite techniques tend to naturally result in greater in-plane TC than in the cross-plane direction, as can be seen in most studies that measure in both directions \cite{Song2015, Kim2016a}. In very thin composites of hundreds of $\mu$m in thickness, often referred to as ``paper TIMs", the in-plane TC can be greater than in the cross-plane direction by orders of magnitude due to the in-plane orientation of fillers \cite{Zhu2014, Luo2016, Kumar2016, Yao2016, Gong2016, Song2016, Song2017, Li2017c, Han2018, Yuan2020}.

A study in 2014 analyzed the thermal performance increase resultant from filling DGEBA with graphene \cite{Prolongo2014}. In this work a modest enhancement of TC was observed relative to what others would find with a similar loading fraction of 10 \it wt. \%\rm graphene fillers of 0.67 W/mK, compared to 0.18 W/mK measured of the pure epoxy. Similar results were obtained previously with a thermal conductivity of 0.65 W/mK with a similar filler, loading level, and polymer matrix \cite{Zhou2013}. In each of these instances, the lateral dimensions were relatively small, as small as 3 $\mu m$, requiring thermal dissipation to often traverse through the highly-insulating matrix. Additionally, graphene intrinsic TC diminishes with reducing lateral size even if larger than the grey phonon mean free path of $\approx$750 nm \cite{Balandin2011, Xu2014}. Since functionalization can aid in the thermal coupling between graphene and matrix, if small flakes are used the benefit of functionalization is more pronounced. Study has been done that directly examined the benefit of functionalization versus graphene size \cite{Shen2016a}. It was determined that functionalization can inhibit composite TC by harming large graphene flake intrinsic TC, establishing a critical flake size at which point any larger flakes would result in composites harmed by the process.

Epoxy polymer TIMs have been crafted and cured directly into an ASTM D5470-inspired copper interface for testing \cite{Park2015}. An interfacial resistance of 3.2 and 4.3 mm$^{2}$K/W for 5 and 10 \it vol. \%\rm, respectively, was measured at 330 K. The TC of each sample was measured to be 2.8 and 3.9 W/mK. These results highlight the need to consider the potential increases in $R_{C1}$ and $R_{C2}$ that an increasing viscosity resultant from an increase in filler level could cause. This outcome of the superior thermally conductive composite having a greater interfacial resistance was observed elsewhere in a polyolefin polymer matrix and was attributed to its mechanical properties \cite{Cui2015a}.

Using graphene derived from Chemical Vapor Deposition (CVD) and subsequently exfoliated, a method that produces graphene of greater quality than that from the reduction of graphene oxide, a TC of 4.9 W/mK was achieved with a 30 \it wt. \%\rm loading in an epoxy resin \cite{Tang2015}. Additionally examined in this study was the TC at different temperatures. There is a reduction of performance at higher temperature as one would expect, but the extent of the reduction proved to be modest, showing positive signs for thermal stability. In a similar work, an epoxy composite with 8 \it wt. \%\rm of graphene achieved a 627\% improvement in TC, resulting in 1.18 W/mK \cite{Wang2015c}. The performance of composites based on these constituent materials can vary substantially from researcher to researcher, displaying the great many influencing parameters that determine their properties. At a similar graphene loading of 8 \it wt. \%\rm, another study reported a TC of $\approx$0.5 W/mK in an epoxy composite \cite{Moriche2016}. These factors that can alter composite performance can range from being intentional and knowable to being difficult to identify.

Generally, graphene without defects is desirable because its TC reduces with increased defect density. However, in a non-equilibrium molecular dynamics simulation, a mechanism for \it increased \rm TC in a liquid n-octane and graphene composite was established with increasing vacancy defects \cite{Liu2015}. Upon introducing vacancy defects to graphene at concentrations up to 8\%, the thermal conductance of the composites is increased because the graphene fillers become more structurally flexible, with a corresponding decrease in its in-plane and out-of-plane phonon frequency. This reduction in out-of-plane vibrational frequency aids in the thermal coupling of the graphene and polymer. This highlights the need to take holistic considerations when designing a composite as opposed to what is traditionally good for an individual component of the composite. Viewing this and other works suggest that defect-based enhancements depend on the type of defect and polymer type \cite{Wang2015a}.

\begin{figure}
	\centering
	\includegraphics[scale=0.95]{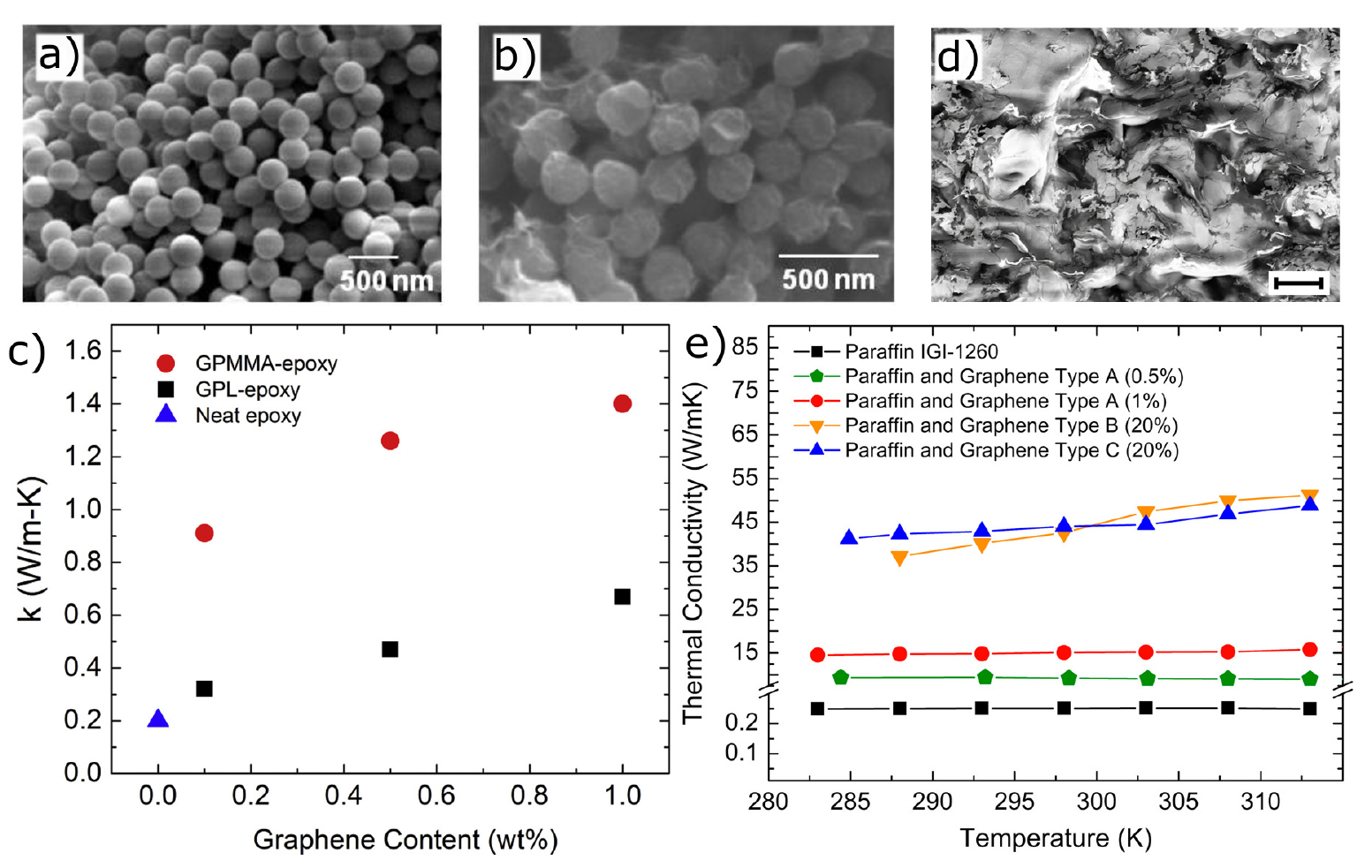}
	\caption{In this study, graphene was grafted onto PMMA spheres to provide structure to the graphene. a) SEM image of pure PMMA spheres. b) PMMA spheres with $16 wt. \%$ graphene c) TC results of graphene attached to PMMA spheres (GPMMA) as red dots and graphene without PMMA attachment (GPL) as black squares. d) SEM micrograph of graphene and phase-change material. e) TC performance of graphene-enhanced Paraffin over realistic battery temperatures. Panels adapted with permission from: \textbf{a, b, c,} ref \protect{\cite{Eksik2016}} and \textbf{d, e,} ref \protect{\cite{Goli2014}}, copyrights 2015 and 2016 Elsevier.}
\end{figure}

An interesting and relatively recent strategy has been to attach graphene to another larger material to achieve a desired larger-scale placement and orientation. This technique was used to make PMMA balls coated with graphene that were then used to fill an epoxy \cite{Eksik2016}. SEM micrographs of varying graphene loadings and magnifications are in Figure (6a-b). Using this technique, the researchers achieved $\approx$ 1.4 W/mK at 1 \it wt. \%\rm, versus only $\approx$ 0.6 W/mK of equivalent loading graphene without PMMA grafting, shown in Figure (6c). A similar idea was applied to attach reduced graphene oxide to thermoplastic polyurethane balls and then hot-press mold the balls together, achieving a TC of 0.8 W/mK at 1.04 \it wt. \%\rm \cite{Li2017a}.

Graphene fillers have been applied to phase change materials, often used in TIMs and thermal energy storage \cite{Peng2004}. An aerogel material's TC has been increased from 0.18 W/mK to 2.64 W/mK with the inclusion of approximately 20 \it vol. \%\rm of graphene oxide \cite{Zhong2013}. The phase-changing polymer icosane's TC was enhanced by a factor of 400\% to $\approx$2.1 W/mK through the inclusion of 10 \it wt. \%\rm of graphene \cite{Fang2013}. These results will allow for better temperature uniformity within each phase-changing polymer due to the enhanced heat flow characteristics with important implications in the ever-more-important lithium battery field \cite{Goli2014}. Figure (6d) an SEM micrograph of a prepared graphene and paraffin composite is shown. In Figure (6e), TC of different graphene-enhanced composites for realistic battery temperatures are presented with $>$45 W/mK performance at slightly above RT. In a lauric acid phase change material a TC enhancement of 230 \% was seen with as little as 1 \it vol. \%\rm \cite{Harish2015}.

\section{Percolation}

As mentioned previously and illustrated in Figure (2), when a composite is loaded past a critical level there can be precipitous increase in conductive ability, whether it be electrical or thermal. This is the case because as the concentration of conductive filler particle increases eventually full pathways from filler to filler forms to allow large-scale low resistance network through the composite. Electrical percolation of composites employing electrically conductive fillers such as metals or carbon allotropes is very strongly supported by research \cite{Biercuk2002, Martin2004, Stankovich2006, Pang2010, Zhang2010a, Potts2011}. The EC of composites are well described by a power law, $\sigma \approx (f - f_{E})^{t}$, where $\sigma$ is the EC, $f$ is the filler volume fraction, $f_{E}$ is the percolation threshold loading level, and $t$ is the critical exponent.

The exact nature and efficacy of thermal percolation in composites was up until recently not considered a settled issue in science \cite{Bujard1988, Choi2001, Shenogina2005, Ding2006, Bonnet2007, Zheng2012, Gu2014, Shtein2015b, Li2017b}. It is clear that the change of composite thermal properties resultant from percolation is more modest than that of EC, which can span over ten orders of magnitude, strongly depending on the matrix and fillers used \cite{Lewis2019}. The less of obvious observable signs of thermal percolation relative to electrical percolation is often attributed to the simple fact that the span of available materials' TC is far more constrained than in the case of EC. The dynamic range of TC -- a total ratio of $K_{f}/K_{m} \approx 10^{5}$ -- in materials that one could use in practical applications is much lower than that of EC -- a total ratio of $\sigma_{f}/\sigma_{m} \approx 10^{15}$, resulting in effectively no polymer electrical conduction while still providing some thermal conduction \cite{Shenogina2005, Kargar2018}. Since the ratio of $K_{f}/K_{m}$ is often ten orders of magnitude less than $\sigma_{f}/\sigma_{m}$, the TC enhancement at the percolation threshold is less precipitous as EC enhancement at its respective percolation threshold.

More recent works have more conclusively shown the onset of a thermal percolation in graphene and \it h\rm -BN composites \cite{Shtein2015b, Kargar2018}. Figure (7a) and (7b) shows TC performance of graphene and \it h\rm -BN showing superlinear TC enhancement after a certain filler loading fraction -- the percolation threshold \cite{Bujard1988, Bonnet2007, Zhang2010b, Wattanakul2011, Kim2011, Zheng2012,Shtein2015b,Gu2014}. The thermal percolations were observed at about 30 \it vol. \%\rm in the graphene composites and 23 \it vol. \%\rm in the \it h\rm -BN composites. The enhancement of TC as the loading fraction is increased was fit to Maxwell-Garnett, Agari, and finally with fantastic agreement, the semi-empirical Lewis-Nielsen model \cite{Garnett1906, Nielsen1973, Nielsen1974, Agari1986, Devpura2000}. This specific behavior is somewhat different to a previous study into graphene composite percolation in which pre-percolation behavior was found to match Nans' model and post-percolation matched the adjusted critical power law \cite{Nan2004, Shtein2015b}. The Lewis-Nielsen TC model is

\begin{equation}
\frac{K}{K_{m}} = \frac{1 + ABf}{1 - B\Psi f}
\end{equation}

where $A$ is equal to $k_{E} - 1$ where $k_{E}$ is the generalized Einstein coefficient, $B = (K_{f}/K_{m} - 1)/(K_{f}/K_{m} + A)$, and $\Psi = 1 + ((1 - \phi_{m})/\phi_{m}^{2})f$ where $\phi_{m}$ is the maximum packing fraction \cite{Pietrak2014}. The values of parameters $A$ and $\phi_{m}$ are unknown for quasi-2D fillers like graphene and \it h\rm -BN and were treated as fitting parameters. 

It was found that loading beyond the thermal percolation threshold placed considerable importance on the cross-plane TC of the graphene fillers because thermal transport in this direction facilitated the passing on of heat from one flake to the next in the percolation network. The graphene composites exhibited consistently higher TC than their \it h\rm -BN counterparts. This fact is due to the superiority of graphene intrinsic TC relative to that of f\it h\rm -BN, at a still impressive experimentally-determined TC of $\approx$230 W/mK to $\approx$480 W/mK at RT and up to $\approx$1000 W/mK when determined theoretically \cite{Sichel1976, Sevik2011, Lindsay2012, Sevik2012, Jo2013, Zhou2014, Wang2016b}.

\begin{figure}
	\centering
	\includegraphics[scale=0.42]{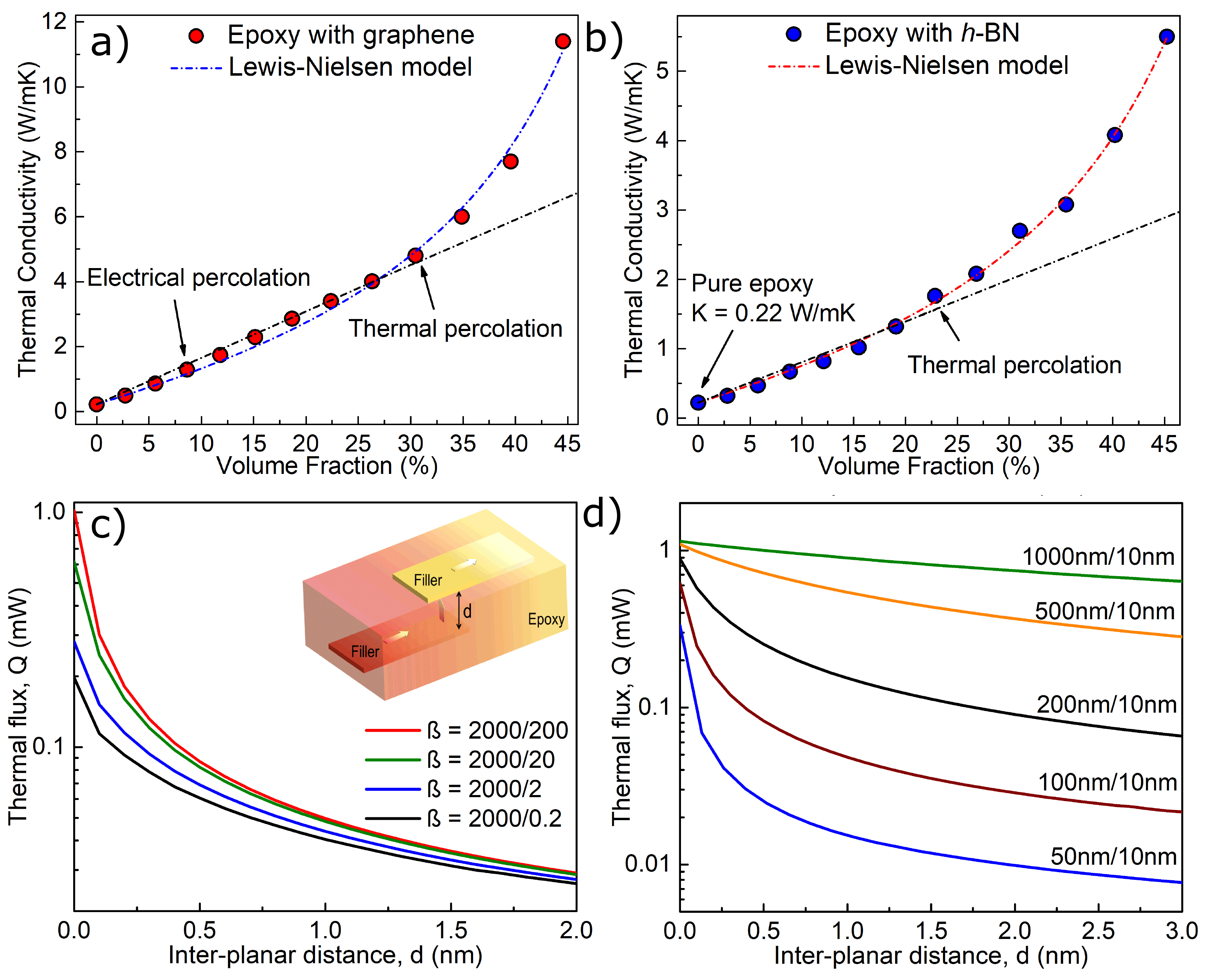}
	\caption{Thermal conductivity versus volume fraction with linear and Lewis-Nielsen trend lines for comparison. a) graphene composites. b) h-BN composites. The superior TC of graphene composites \textit{h}-BN composites is attributed to the superior intrinsic TC of graphene. c) Thermal flux versus distance between flakes shown in schematic. Each line corresponds to a simulation result with a different cross-plane TC, varying from 0.2 to 200 W/mK. d) Same plot for varying flake lateral dimensions. Adapted with permission from ref \protect{\cite{Kargar2018}}. Copyrights 2018 American Chemical Society.}
\end{figure}

Using the Lewis-Nielsen model, a surprisingly low \it apparent \rm TC of $\approx$ 37 W/mK was determined for the graphene materials used inside of the TIM. This lower-than-expected TC was attributed to the unexpectedly important impact of filler cross-plane TC to the overall thermal transport. If the composite is filled past its percolation threshold, much of its heat will be transporting from one flake to another laying on top of it, forcing transport in the cross-plane direction. TC in this direction can be 2 orders of magnitude less than in the in-plane direction. It is also possible that the matrix and filler defects can induce TC-harming phonon scattering, however the amount of scattering necessary to alone explain the low apparent TC of graphene seems less likely. The effect of microscopic contact resistance -- Kapitza resistance -- is likely a contributor and could be greatly diminished in future works with functionalization processes \cite{Hopkins2012, Foley2015, Walton2017}.

Figure (7c) shows a comparison of thermal transport for different composite parameters from a finite element heat diffusion numerical simulation. The subset schematic in subfigure (7c) shows a quasi-2D filler within an epoxy matrix. This filler has a heat applied and that heat is transported via diffusion away from the schematic's exposed face towards the end of the flake, then the heat traverses primarily vertically through the epoxy, across a distance d, and into another filler. Plotted in (7c) is the thermal flux of flakes with high-quality graphite's in-plane TC of 2,000 W/mK and various cross-plane TCs with different distances between the adjacent flakes. Evident from the plot is the considerable importance of the overall thermal flux, amounting to a factor of $\approx$5, on the cross-plane thermal conductivity when the fillers are making contact, such as in the thermal percolative state. In figure (7d) the total thermal flux versus distance between flakes is considered for varying flake lateral sizes and fixed thicknesses. The importance of large flakes below the percolation threshold, and thus large inter-planar distance, is clear and is due to the opening of long, low resistance pathways and the reduction of reliance on the comparatively low cross-plane TC.

Recently, a new composite TC differential equation model was reported that agrees well with this work \cite{Drozdov2019}. The model is written as,

\begin{equation}
\frac{\mathrm{d}X}{\mathrm{d}\phi} = \frac{1}{1 - \phi} \Bigg[ \frac{R_{1} (1-\Lambda)}{3} + \frac{B\Lambda(R_{2}-X)X}{R_{2}+(B-1)X} \Bigg]
\end{equation}

where $X$ is the ratio of the final composite thermal TC to the pure matrix TC, $\phi$ is the filler volume fraction, $R_{1}$ and $R_{2}$ are the ratios of the filler effective TCs to that of the matrix, $\Lambda$ is the volume fraction of particles that are in tight clusters resultant from imperfect mixtures, and $B$ characterizes how particles and their clusters deviate from a spherical shape. This model is aware of thermal boundary resistances, percolative networks, and imperfect mixture agglomerations. 

\section{High Loading Non-Curing Graphene Thermal Interface Materials}

Cured, solid form TIMs receive a more attention in research possibly because of the ease of working with them relative to non-curing forms, in addition to their direct comparison to chip encapsulation materials. However, a more representative comparison between the TIMs used in VLSI package and heat sink junctions can be made in studies of non-curing, at least semi-fluid TIMs, despite their relative difficulty to work with. It is common for non-curing TIMs to be out-performed by curing TIMs, all other things being held equal including polymer base TC. Current commercial non-curing TIMs currently have a bulk thermal conductivity range of 0.5 to 7 W/mK and are needed to reach 20 to 25 w/mK to allow for next-generation devices \cite{Gwinn2003, BarCohen2015}. 

Research into graphene-enhanced non-curing TIMs was up until recently exclusively studied using commercial TIMs as the matrix. These matrix materials typically start at a relatively high viscosity primarily due to having their own filler materials already incorporated, leaving little headroom in which one may add additional fillers. In spite of this, addition of small quantities of graphene into these materials has shown impressive TC improvements \cite{Shahil2012, Goyal2012, Kargar2014, Hansson2016}. The presence of the commercial TIMs' undisclosed filler materials makes detailed analysis of the observed behavior difficult.

This group worked on a graphene-based non-curing TIM with a simple mineral oil base matrix for both greater insights into material properties and more room with respect to viscosity to further load with graphene \cite{Naghibi2020}. The $\approx$15 $\mu$m lateral dimension graphene were mixed in with the mineral oil in addition to acetone to prevent agglomeration \cite{Ma2019, Hou2019}. After mixing, the acetone was removed from the mixture by exposure to 70 $^{\circ}$C for $\approx$2 hours in a furnace. It was suspected that the incorporation of acetone in the mixing process helped preserve the filler quality.

Using the popular ASTM-D5470 steady-state technique, the junction thermal resistance and TCs of these composites were characterized between two parallel plates. The thermal resistances of the composites between the two plates at different distances and composite concentrations are shown in Figure (8a). The inverse of the slope for every fitted line for each composite corresponds to its TC. The y-intercept of this fitted line is the sum of $R_{C1}$ and $R_{C2}$ in equation 1, which are equivalent to one another given the top and bottom junctions were identical. The reduction of the slope of the composites' fitted lines with increasing graphene content indicates the steady increase of bulk TC for increasing filler loadings. As previously discussed, the increasing importance of TIM TC in real-world BLTs of 300 $\mu$m is clearly presented by these findings.

\begin{figure}
	\centering
	\includegraphics[scale=.35]{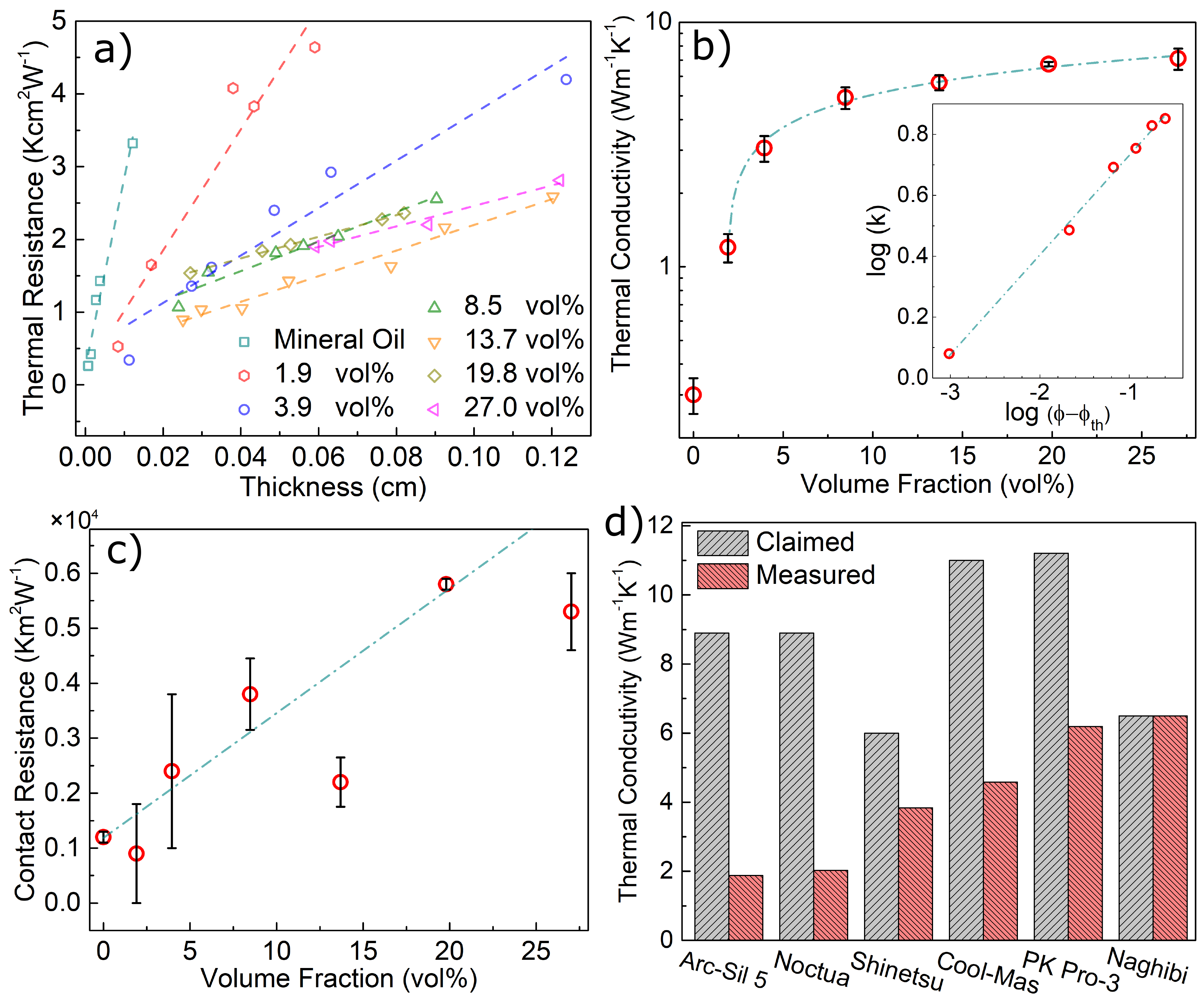}
	\caption{a) Thermal resistance per unit area versus BLT. b) TC as a function of volume fraction determined from the inverse of the slopes in a. c) Contact resistance versus volume fraction with behavior dominated by the role of viscosity relating the two parameters. d) Comparison of graphene TIMs studied with claims of TIM vendors studied with ASTM-D5470. Adapted with permission from ref \protect{\cite{Naghibi2020}}. Copyright 2020 John Wiley \& Sons, Inc.}
\end{figure}

Using the inverses of slopes from Figure (8a), Figure (8b) shows the derived TCs of the tested composites. The error bars are convey the from errors in the linear regression. A sharp increase of TC, from 0.3 W/mK to 1.2 W/mK, is seen after applying a relatively low loading of 1.9 $vol. \%$ indicating an early onset of thermal percolation, followed by the beginning of saturation behavior at 8.5 $vol. \%$. This behavior is well matched with a power scaling law, $K_{TIM} = A (\phi - \phi_{th})^{p}$, where $A$ is a fitting parameter related to the effective TC with consideration to boundary resistance, $\phi_{th}$ is the percolation threshold, and $p$ is the universal exponent. TC saturation in non-curing TIMs has been observed previously though is generally absent in works into curing composites \cite{Prasher2003, Zhang2015, Mu2016}. The saturation of TC is attributed to an increase of filler interface resistance as the concentration of graphene increases as a specific interaction between the filler and this individual polymer matrix \cite{Evans2008}.

Figure (8c) shows the contact resistance of the tested composites, with increasing contact resistance for increasing loading fraction. Assuming the bulk TC of the composite is negligible in comparison to that of the mating faces in the junction, the contact resistance can be described by the following semi-empirical model:

\begin{equation}
R^{\prime\prime}_{C_{1}+C_{2}} = 2 R^{\prime\prime}_{C} = c \Bigg(\frac{\zeta}{k_{TIM}}\Bigg) \Bigg(\frac{G}{P}\Bigg)^{n}
\end{equation}
where $G = \sqrt{G^{\prime 2} + G^{\prime \prime 2}}$. $G^{\prime}$ and $G^{\prime \prime}$ are the storage and loss modulus of the TIMs, $P$ is the applied pressure of atmosphere in this case, $\zeta$ is the average roughness of the two identical surfaces, and $c$ and $n$ are empirical coefficients \cite{Prasher2003}. Predicting the thermal contact resistance with any accuracy from successive experiments at constant pressure is challenging given that the two remaining parameters -- $k_{TIM}$ and $G$ -- are affected by graphene loading and oppose one another in the determination of $R^{\prime\prime}_{C}$. This equation exposes that in TIMs well-described by it there is an optimum filler loading in which $k_{TIM}$ may be substantially enhanced with little increase in $R^{\prime\prime}_{C}$.

The bulk TC of the present 19.8 $vol. \%$ graphene TIM is compared with high end commercial TIM products in Figure (8d). Industry self-reports TCs higher than 11 w/mK but do not disclose the technique used to arrive at those values. Here we present all of the TIM TCs measured with the ASTM-D5470 technique compared with the values reported by the manufacturer's. The 19.8 $vol. \%$ graphene TIM performs better than all tested commercial TIMs. The closest performing TIM -- PK Pro-3 -- uses $\approx$90 $wt. \%$ of Aluminum and Zinc Oxide fillers, over two times the loading level of the graphene TIM compared.

These TIMs have been applied to solar cells to study the reduction of performance resultant from operating at elevated temperatures \cite{Mahadevan2019}. The poorer performance appears as a decrease in the voltage across the cell's two terminals. For every increase in operating temperature in degree Celsius above 40 $^{\circ}$C there is an efficiency loss of 0.35\% to 0.5\% \cite{Brinkworth1997, Radziemska2003}. Silicon-based solar cells are known to reach temperatures up to 65 $^{\circ}$C, corresponding to up to a 12.5\% decrease in efficiency.

\begin{figure}
	\centering
	\includegraphics[scale=1.4]{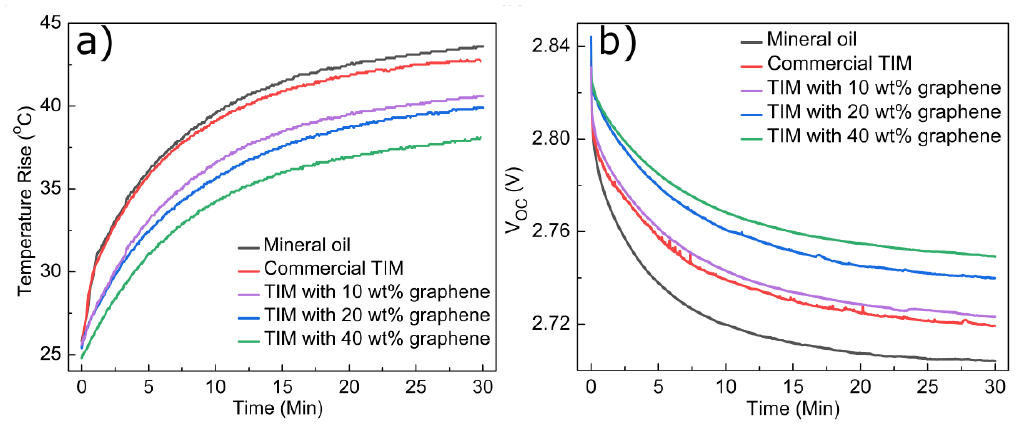}
	\caption{a) Temperature of a solar cell over time under 70x natural solar illumination with different TIMs applied between it and a heat sink. b) The corresponding open-circuit Voltage of the solar cell resultant from the device temperature. Adapted with permission from ref \protect{\cite{Mahadevan2019}}. Published under the CC license by MDPI.}
\end{figure}

It is common practice in solar cell research to analyze its performance under simulated sun light and at greater-than-natural illumination to among other reasons, provide the heat elevate the device in test above RT \cite{Lammert1977, Han2011}. In this study, a solar cell was fixed to a heat sink with different TIMs applied between and was illuminated with 70x and 200x natural solar illumination levels, the former being considered at present. Figure (9a) shows the temperature change of a solar cell over time with different TIMs. It is evident that when the solar cell had the higher graphene concentration TIMs the temperature that it reached remained lower, showing a better thermal coupling to its heat sink. Figure (9b) shows the corresponding open-circuit voltages -- a common photovoltaic metric of efficiency -- that displays the increased efficiency gained for maintaining a lower operating temperature.

\section{Hybridization and Control of Electrical Conductivity}

Researchers have long noted a beneficial TC performance of composites that employ multiple types of fillers, a filling strategy known as hybridization or binary, tertiary, etc. filling \cite{Wang2003, Sim2005, Yu2008, Kemaloglu2010, Li2010, Yang2010, Zhou2010, Pak2012, Teng2012, Liu2017, Wang2018b,  Dmitriev2020}. This synergistic effect is seen when including multiple filler materials at a certain constituent ratio can achieve a greater TC enhancement than with either individual filler at identical overall loading level. This effect arises from the differing morphology of the two filler materials and how they can aid one another. Despite the phenomenal intrinsic TC of graphene that one could reasonably expect to overpower any potential synergistic effect, it has been widely reported in graphene composites \cite{Shahil2012,Gao2015, Cui2015b, Shao2016, Zhang2019}. This benefit occurs due to a second filler's ability to prevent graphene agglomeration in a composite and its ability to bridge gaps between graphene flakes that would otherwise force heat transport through the resistive polymer.

Due to the frequent desire for high TC but low electrical conductivity TIMs, hybridization is a promising way to leverage the extremely high TC graphene fillers while controlling the resulting composite electrical conductivity that they cause. It was shown previously that a hybrid composite of very disparate geometries of graphene flakes and boron nitride nanoparticles could achieve synergy and a suppression of composite electrical conductivity \cite{Shtein2015a}. In this work the electrically conductive graphene flakes were effectively isolated from one another by the smaller electrically insulating boron nitride materials fitting between them, allowing thermal but not electrical conduction. This can be seen in Figure (10a) and (10b) in a SEM image and a schematic showing smaller, red boron nitride fitting between blue graphene flakes. The superiority of composites' TC with a hybridization of filler material along with a reduction in EC can be seen in Figure (10c).

\begin{figure}
	\centering
	\includegraphics[scale=.75]{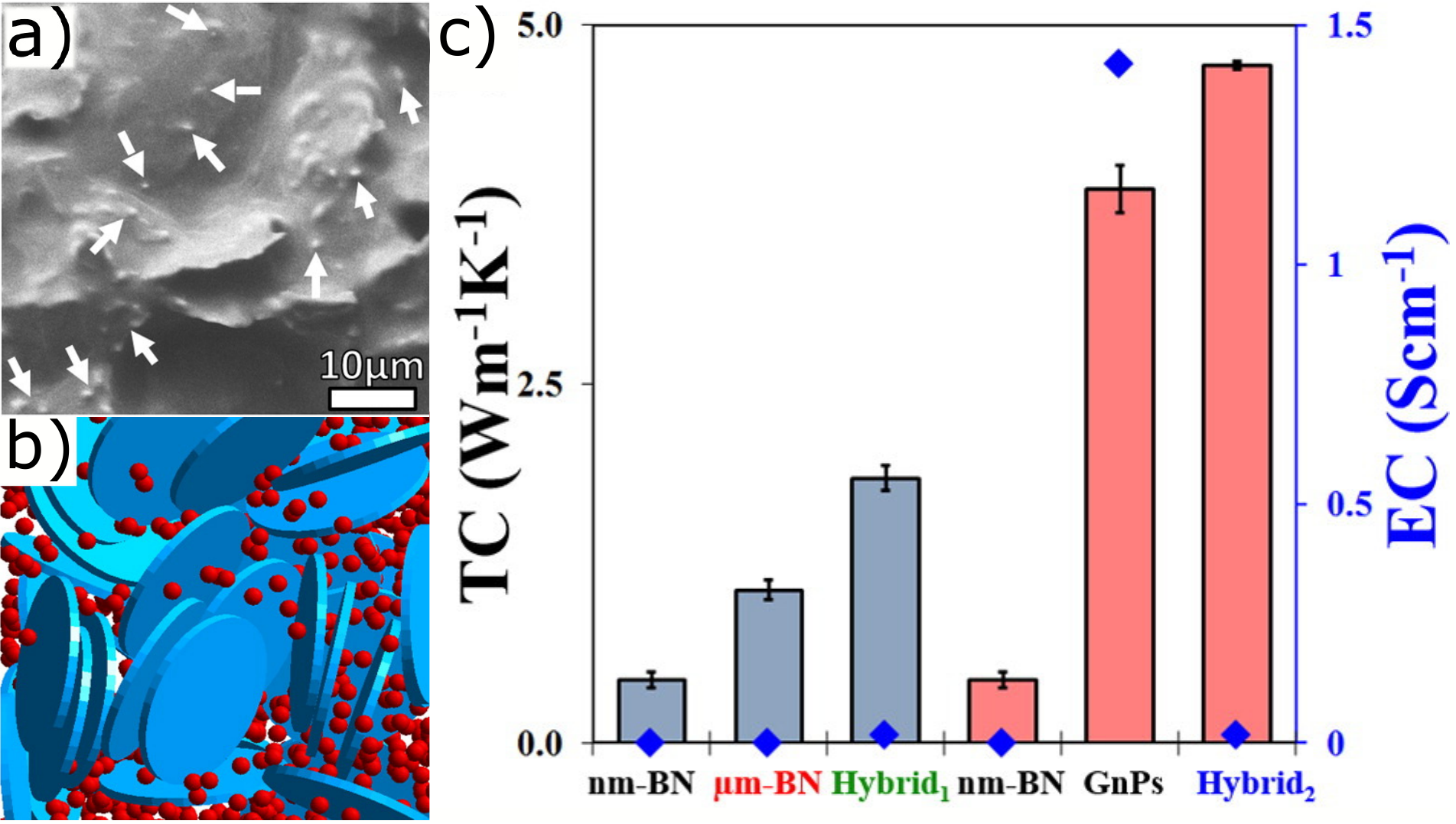}
	\caption{a) Schematic showing mixed graphene flakes and smaller boron nitride particles. b) SEM image of an epoxy composite with arrows pointing out boron nitride particles. c) Plot of TC and EC of composites with total filler loading of 17 $vol. \%$ composed. ``nm-BN" corresponds to composites filled with boron nitride of 200 nanometers in lateral dimensions. ``$\mu$m-BN" corresponds to composites of boron nitride of approximately 40 microns. ``Hybrid 1" is $15 vol. \%$ of $\mu$m-BN and 2 $vol. \%$ of nm-BN. ``Hybrid 2" is $16 vol. \%$ of GnP and 1 $vol. \%$ of nm-BN. Note the increase of TC relative to the composite of pure GnPs, which is certainly a result of synergy, as well as the sharp reduction of EC. Adapted with permission from ref \protect{\cite{Shtein2015b}}. Copyright 2015 American Chemical Society.}
\end{figure}

This research group prepared hybrid composites of graphene and \it h\rm -BN flakes of similar geometries to investigate both whether one can achieve a more finely-tuned control on electrical conductivity and as a contrapositive verification of each filler's dissimilar geometries in producing a synergistic effect \cite{Lewis2019}. Figure (11a) shows a schematic of the use of hybrid fillers to selectively control composite electrical conductivity while preserving useful TC. The graphene and \it h\rm -BN flakes used both had thicknesses up to 12 nm and lateral dimensions up to 8 $\mu$m. It was hypothesized that if the two materials were of comparable geometries then they would be less effective at isolating one another than had been observed before. 

\begin{figure}
	\centering
	\includegraphics[scale=0.75]{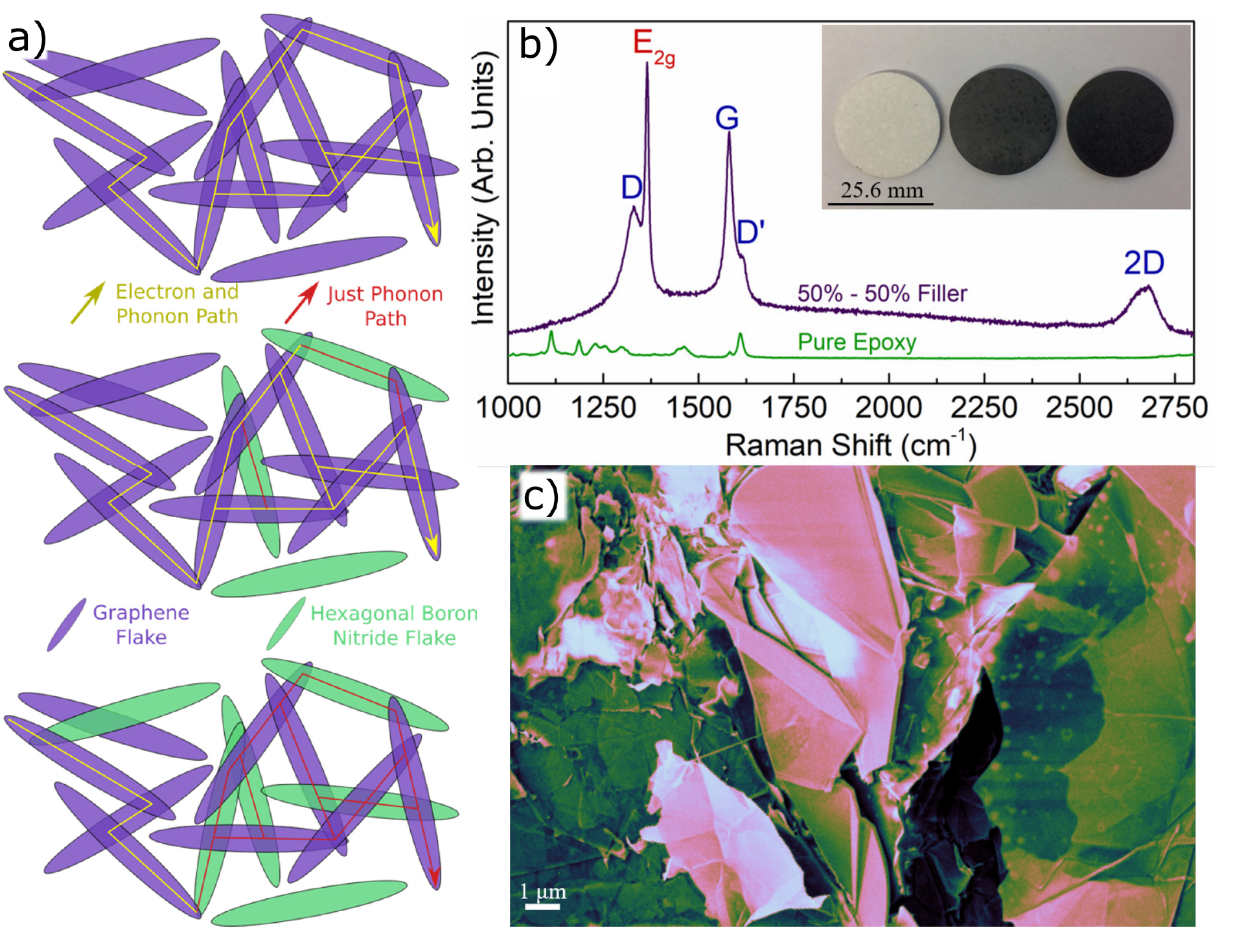}
	\caption{a) Raman spectrum of half graphene and half \it h\rm -BN composite with total loading of 44 \it vol. \%\rm . b) pseudo-colored SEM image of a fractured surface of a 13 \it vol. \%\rm of graphene and \it h\rm -BN each composite. c) Top schematic shows a pure graphene composite in which electrons and phonons freely move throughout. Middle schematic shows some boron nitride flakes of similar geometries thrown in which create paths in that only easily transmit phonons, but not electrons, reducing overall composite electrical conductivity. Bottom schematic shows a concentration of boron nitride flakes where entire electrical percolation networks have been disrupted. Panels \textbf{b, c} adapted with permission from ref \protect{\cite{Lewis2019}}. Copyright 2019 IOP Publishing.}
\end{figure}

Figure (11b) shows Raman signatures of a 44 \it vol. \%\rm composite with 50\% constituent fraction of graphene and 50\% of \it h\rm -BN. Characteristic peaks of graphene and its disorder are present as well as the $E_{2g}$ peak of \it h\rm -BN \cite{Ferrari2006, Ferrari2007, Calizo2007, Cai2017}. The inset image shows selected high loading samples. Figure (11c) shows a pseudo-colorized fractured surface of composite with pink, electrically charging \it h\rm -BN flakes dispersed among green and blue electrically conducting graphene flakes. This image shows at least in one instance the isolation that would not have occurred if \it h\rm -BN flakes were not substituted in for graphene.

These composites' thermal diffusivities were measured using the common laser flash analysis (LFA) technique \cite{Parker1961, Gaal2004}. Using densities determined from Archimedes' principle and heat capacity calculated from the rule of mixtures, thermal conductivity is calculated from the classic relation $K = \alpha \times \rho \times c_{p}$, where $\alpha$ is the thermal diffusivity, $\rho$ is the volumetric mass density, and $c_{p}$ is the specific heat capacity. LFA directly measures $\alpha$, but combining LFA with techniques to determine the other material parameters is an exceedingly popular TC measurement strategy. The heat capacity was calculated using 0.807 J/gK for \it h\rm -BN and 0.72 J/gK for graphite, which only notably deviates from graphene to the ZA phonon dispersion in graphite whose states can be unfilled below 100 K \cite{Dworkin1954, Butland1973, Hone2001, Nika2014, Cocemasov2015}.

\begin{figure}
	\centering
	\includegraphics[scale=0.8]{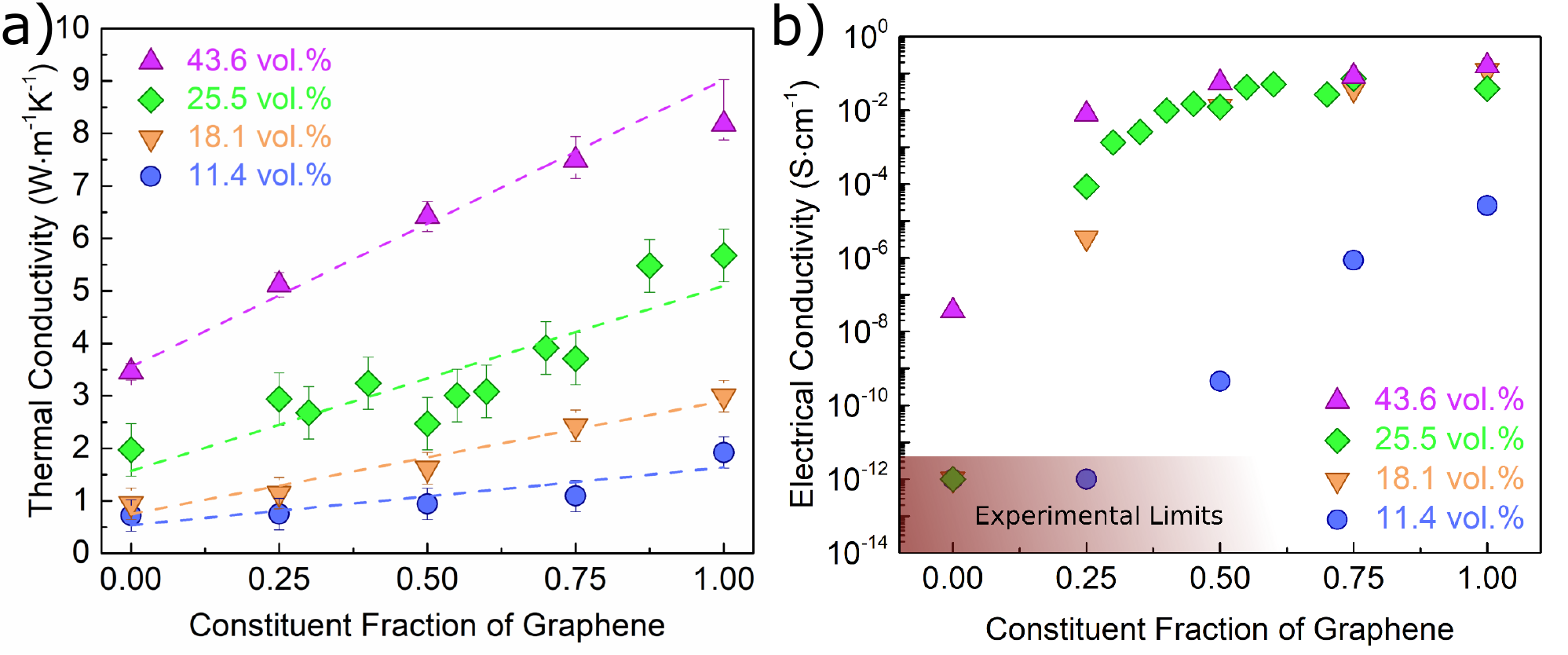}
	\caption{a) Thermal conductivity of each set of total filler level versus the graphene/\it h\rm -BN constituent ratio. Data points at 1.00 on the x-axis correspond to a sample at the stated $vol. \%$ composed of only graphene and matrix. b) Electrical conductivity of the studied composites displaying a power scaling law. Adapted with permission from ref \protect{\cite{Lewis2019}}. Copyright 2019 IOP Publishing.}
\end{figure}

Figure (12a) displays the TC of color-coded composites of 11.4 $vol. \%$, 18.1 $vol. \%$, 25.5 $vol. \%$, and 43.6 $vol. \%$. In all instances as the total filler level is increased the overall TC is enhanced relative to that constituent fraction at a lower total loading. As the constituent fraction of the composites moves to higher levels of graphene (left to right on the x axis), the TC is uniformly enhanced. This result shows that a synergistic enhancement was not observed in these composites. In all tested composites, the superiority of graphene to that of $h$-BN remained the dominant factor. This provides contrapositive verification of the attribution of synergy to dissimilar filler geometries. The increased data scatter in the 25.5 $vol. \%$ is ascribed to that filler loading percentage's proximity to the percolation threshold in composites of this matrix and filler geometry. This would result in some composites stochastically achieving better percolation networks than others, whereas composites above or below this loading are either firmly within or outside of a percolative filling regime. The asymmetric error bar on the 100\% graphene sample at 43.6 $vol. \%$ is attributed to clear error in the measurement of that sample's density resultant from surface bubble formation.

The cross-plane electrical conductivities of the composites were measured by simply painting silver electrodes on opposing faces of the samples and measuring the resistance from the two-probe method in a process that has been done in similar studies \cite{Sandler1999, Wang2008}. Figure (12b) shows the electrical conductivity in the same manner as Figure (12a) with the constituent fraction of graphene on the x axis.

The obtained electrical conductivity results show a range of at least 11 orders of magnitude, though the full range is obscured due to experimental limitations. For all total filler levels, a strong dependence on the constituent fraction of graphene is observed with a power law dependence. At total filler level greater than 11.4 $vol. \%$ composites' electrical conductivities saturate at constituent graphene level of 25\%. 

The lack of a synergistic effect in these composites is supported by both the thermal and electrical conductivity results. The thermal conductivity shows a linear trend when altering one filler concentration over to the other. The accepted explanation for synergistic enhancement of composite TC is that a smaller filler could fit between larger fillers and provide a sort of thermal bridge between the two longer-range fillers. In effect, this would decrease the percentage of distance that must be traveled through the highly insulating polymer matrix material along any pathway. Because the two used filler materials had similar geometries any additional enhancement resultant from one filler efficiently fitting between another was not observed. It has been shown previously that leveraging this dissimilar geometry in $h$-BN and graphene can have dramatic affects on the electrical conductivity of the composite, which was not seen at present \cite{Shtein2015b}. The sharp reduction of electrical conductivity after the introduction of $h$-BN in this work was attributed to the smaller $h$-BN fitting between the electrically conductive graphene material and isolating them, preventing long-range electrical percolation. The results of these two works are in great agreement with one another and help to unequivocally explain behaviors in these hybrid composites.

Another work specifically on composites with sheets of $h$-BN of $\approx 250$ nm and sheets of graphene of $\approx 5$ $\mu$m in lateral size also noted TC synergy \cite{Cui2015b}. The authors achieved a TC of 1.31 W/mK in polyamide and 20 $wt. \%$ graphene compared to 0.28 W/mK in pure polyamide. When the authors included a mere 1.5 $wt. \%$ in addition to the graphene they achieved a thermal conductivity of 1.76 W/mK. It is possible that this marked enhancement relative to the previous sample with only graphene is indeed likely due to synergistic enhancement of the two fillers. It is true that in these composites the overall filler loading is increased and could be approaching the percolation regime of thermal performance, though filler loading is lower than where this group has typically seen the onset of percolation.

Hybrid composites have been interestingly investigated with graphene and alumina spheres of multiple diameters (5 $\mu$m and 0.7 $\mu$m), effectively using three loading materials in a silicone oil matrix \cite{Yu2015b}. The different size alumina fillers are varied as a parameter to achieve higher packing density at the expense of larger fillers to allow larger unobstructed thermal pathways \cite{Cumberland1987}. In a composite only composed of the two alumina sphere sizes, an optimal synergistic ratio of 15 $vol. \%$ of smaller alumina and 45 $vol. \%$ of larger alumina was observed. This optimization between the concentrations of two filler types is the defining characteristic of the synergy mechanism. Raising the total concentration at fixed constituent ratio of the alumina composite from 60 $vol. \%$ to 63 $vol. \%$ results in an increase of 0.49 W/mK, and adding just 1 $wt. \%$ of graphene results in a further 0.75 W/mK improvement up to a total of $\approx$ 3.5 W/mK. Similarly, another study found that epoxy filled with 80 $wt. \%$ had a TC of 0.8 W/mK while substituting the last 7 $wt. \%$ for graphene achieved a TC of 1.8 W/mK \cite{Guan2016}. 

\begin{figure}
	\centering
	\includegraphics[scale=0.75]{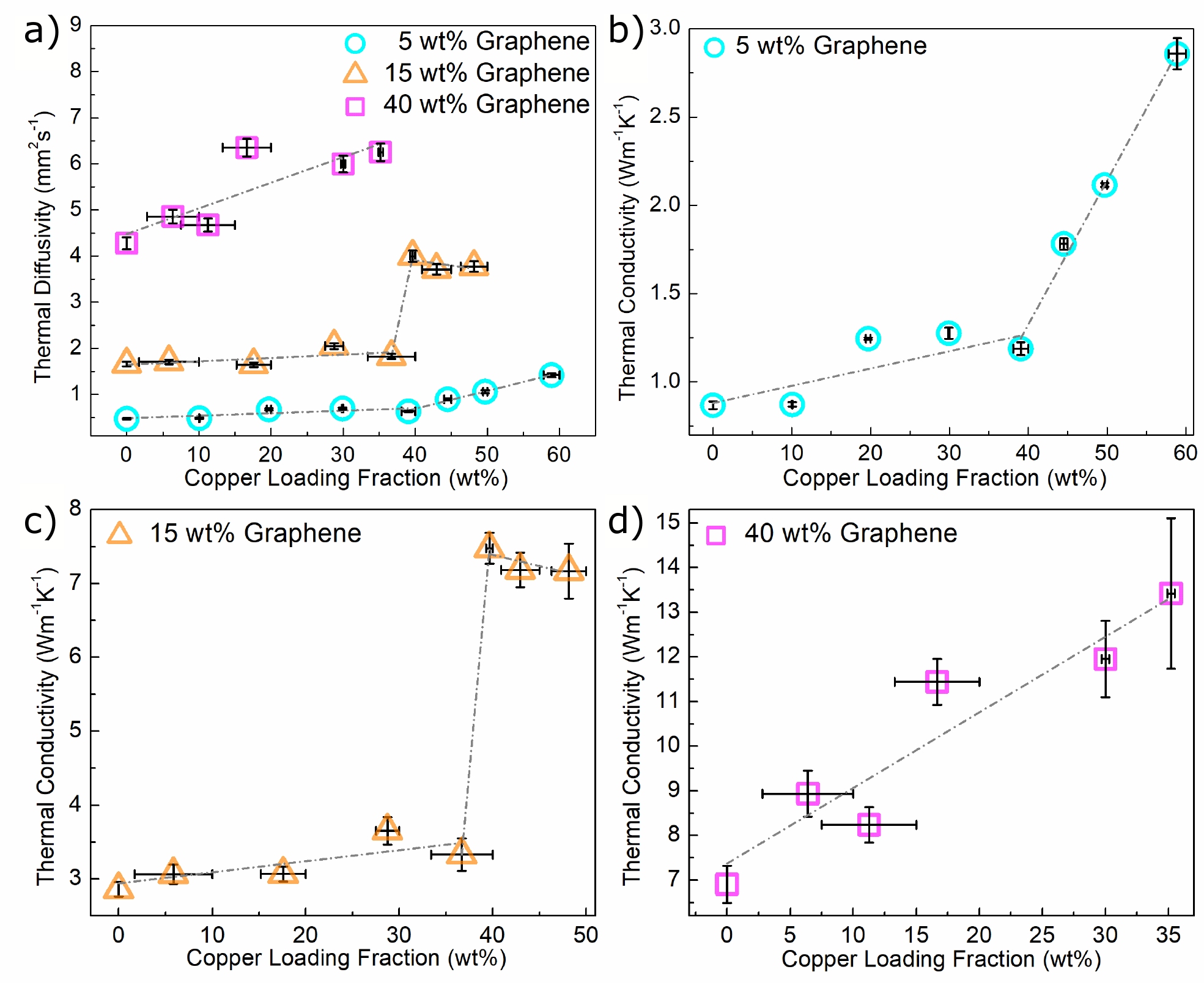}
	\caption{a) Thermal diffusivity of 5 $wt. \%$, 15 $wt. \%$, and 40$wt. \%$ graphene TIMs with increasing copper loading as binary TIMs. Subfigures (b), (c), and (d) are TCs calculated for each graphene concentration plotted in subfigure a. Adapted with permission from ref \protect{\cite{Barani2020}}. Copyright 2019 John Wiley \& Sons, Inc.}
\end{figure}

Recently, this group published a study of graphene and copper nanoparticle hybrid-filled TIMs that exhibited likely synergistic thermal properties \cite{Barani2020}. This work used graphene graphene with lateral dimensions of $\approx$25 $\mu$m and copper spheres with diameters of 40, 100, and 580 nm. Given the copper nanoparticles' conformance to the Wiedemann-Franz Law, it is vitally important to preserve the electrical conductivity of the material to, in turn, preserve the thermal conductivity. To that end, the smallest copper nanoparticle size corresponded to roughly the electron mean free path in copper. Generally, the mean free paths of whatever dominant heat carrier of a considered filler material is a crucial consideration in the minimum size that can still effectively transport heat. In the case of copper nanoparticle fillers, extraordinary care must be taken to prevent rapid and unsafe oxidation that can reduce the thermal conductivity by an order of magnitude \cite{Liu2006}. Figure (13a) shows the thermal diffusivity of 5 $wt. \%$, 15 $wt. \%$, and 40$wt. \%$ graphene TIMs with increasing copper loading as binary TIMs. As expected, the composites that contain a higher load level of graphene fillers have a higher thermal diffusivity. Figures (13b-d) show the calculated TC of each composite. Notably in Figure (13c), a sharp increase in the TC of the 15 $wt. \%$ TIM can be seen between 35 and 40 $wt. \%$. This dramatic enhancement in TC followed by little improvement, even possibly a slight reduction, suggests that a critical optimum of constituent fraction has been reached and moving past it does not further improve performance.

\section{Lifespan reliability and performance}

Along with the associated costs, one of the primary reasons polymeric TIMs receive such preferential usage in industry is due to its lifespan performance versus, for instance, metallic and pad TIMs. It is perceived by the current authors to be a short-coming of TIM research that lifespan performance of novel TIMs is so seldom considered, especially in graphene-based TIMs, likely borne from the time commitment such a study would entail. TIMs are by their very nature applied in very difficult environments and need to maintain performance for as long as possible, very often for the entire lifespan of the device. As the devices are turned on and off, operated in humid environments, and exposed to environmental contaminants their intrinsic material characteristics can alter as well as the morphology of the mating surface in which they are applied. Each of these alterations can lead to catastrophic failure from cracking or being pumped out of the junction as a result of the thermal expansions, contractions, and warping over the course of high and low power device state fluctuations causing wide temperature alterations. Perhaps the largest factor affecting the lifespan performance of TIMs in-junction is the coefficient of thermal expansion mismatches between TIM and junction. The problems that can arise can take the form of cracks, voids, or intrinsic denaturing of the TIM. Figures (14a-c) show acoustic microscopy images of a TIM application that is still in tact, exhibits voids, and has cracks \cite{Gupta2006}. Figures (14d-f) shows corresponding infrared thermography images a few milliseconds after powering the device that shows faster heat spread on the TIM with superior coverage.

\begin{figure}
	\centering
	\includegraphics[scale=0.7]{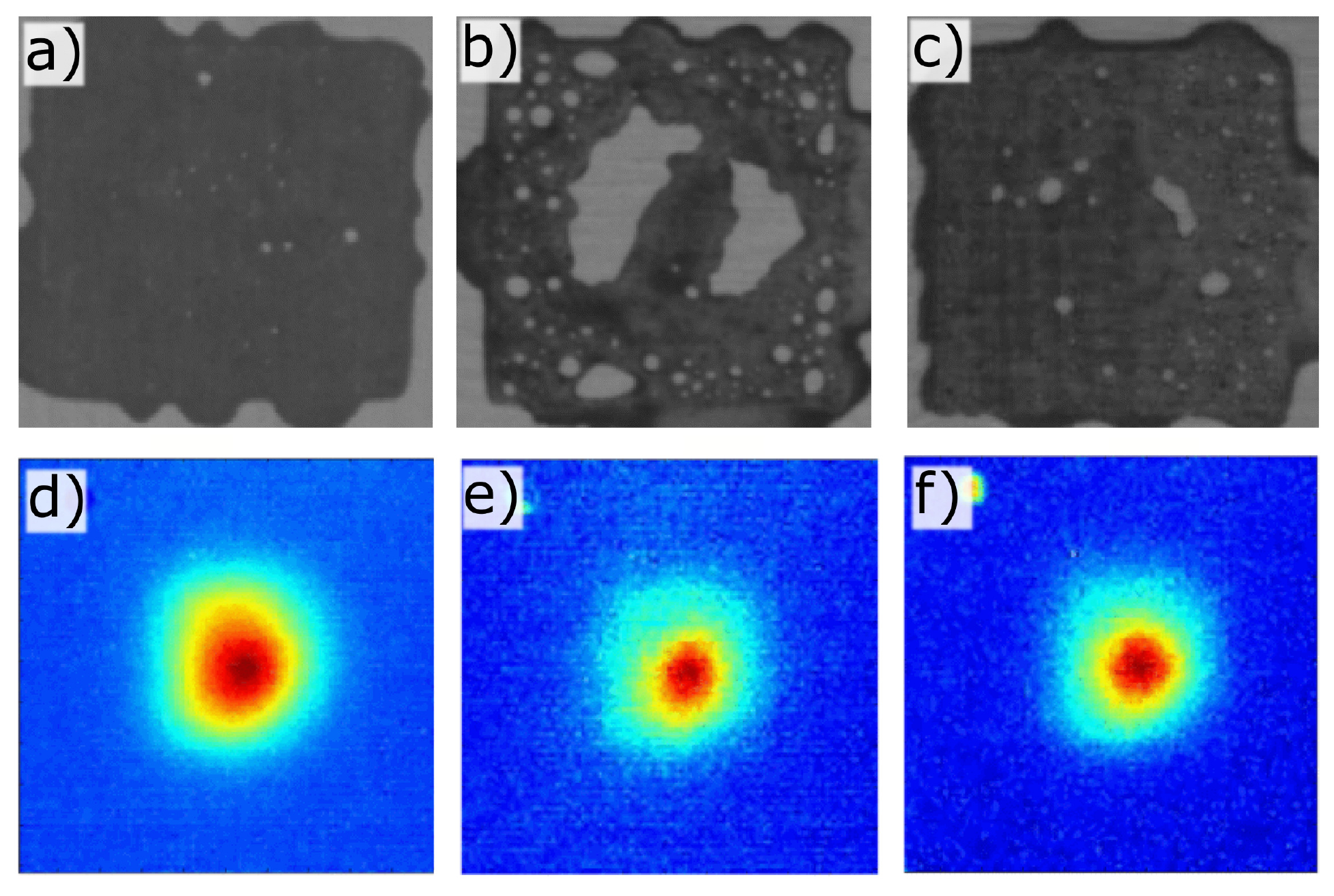}
	\caption{Top rows are scanning acoustic microscope and bottom are IR thermography images taken a few milliseconds after device powering on. a) is a high quality TIM application, b) is a TIM application with substantial voids, and c) is a TIM application with micro-cracks. d), e), and f) are IR thermography images corresponding to a), b), and c), respectively. The quicker spread of temperature in d) than e) and f), evidenced by the wider red and green region demonstrates a quicker heat spread than the other two samples. Adapted with permission from ref \protect{\cite{Gupta2006}}. Copyright 2006 IEEE.}
\end{figure}

Though the fraction of published works that report lifespan performance to total works published in polymeric TIMs is quite low, researchers have considered this often overlooked aspect \cite{Gupta2006, Due2013, Li2017a, Li2020}. The literature on this matter, unfortunately, is quite inconsistent likely due to the lack of a universal standard technique for reliability and the likelihood that any developed standard technique would be unable to provide predictive performance for every individual device application. There are three classes of accelerated aging techniques that most of the experiments conducted into TIM reliability can be categorized within: Elevated temperature storage, temperature cycling, and power cycling \cite{Due2013}.

Elevated temperature storage procedures hold a TIM typically in a junction sandwich at a uniform elevated temperature for an extended period of time. Very importantly, they may or may not employ a high humidity environment to simulate important moisture interactions. The performance of TIMs in this test varies greatly depending on the TIM and junction materials, showing both enhanced and hindered performance over the course of treatment \cite{Luo2000, Dal2004, Ramaswamy2004, Bharatham2005, Gowda2005, Khuu2009, Chen2009b, Paisner2010}. Likewise, a TIM can either experience enhancement from humidity resultant from increased wetting or experience harm the adhesion ability of the polymer matrix \cite{Dal2004, Goel2008}.The lack of consistency in this type of procedure has numerous causes from differences in the procedure, different materials, chemical degradation, and physical form changes.

More representative of realistic TIM conditions is the temperature cycling procedure. In this technique the TIM often inside of an overall junction are cyclically placed in uniform high and low temperature environments. This procedure more closely approximates real-world TIM conditions because of the fact that TIMs operate at a wide range of temperatures. This procedure allows for multiple thermal expansions and contractions to occur, which is an important parameter in TIM pump out and cracking. The results in literature for this procedure are inconsistent as well, with most non-curing TIMs performing better \cite{Gowda2003, Gowda2005, Paisner2010}. It was observed previously that most of these instances of improvement were attributed to a reduction of the BLT and increased wetting, each mechanism not a contributing factor to cured TIMs \cite{Due2013}.

Likely the most representative accelerated aging method is power cycling. In this technique a TIM often with its accompanying junction are cyclically heated from a localized source, resulting in a temperature gradient. This method captures thermal expansion and contraction mechanisms experienced in TIM applications the closest. Non-curing TIMs typically exhibit a reduction in performance between 20\% and 60\%, showing the superiority of this technique in reproducing real world trends \cite{Due2013}. Because the sample is being heated from one side, it is of greater importance that one consider the rate of heating. If the heat were too high in the localized spot that the heater is located then it would increase the effective thermal expansion mismatch in either just the TIM or the entire TIM and junction sandwich. 

This group worked on a power cycled reliability study on graphene-filled epoxy TIMs, without an adjoining junction \cite{Lewis2020}. The decision to not examine the TIM inside a junction sandwich stemmed from a desire to analyze the intrinsic thermal conductivity lifespan performance and. A custom Nichrome wire heating loop between Kapton was fabricated to be used as the localized heating element. As part of a control system, a Type-J thermocouple was fixed to the back of the sample as a feedback to inform how much electrical power to supply the heating coil. In the Python programming language, an elementary machine learning algorithm determined the amount of power that was needed to supply to the coil to achieve the desired temperature range without any assumptions of material properties, then it ran unattended with intermittent re-calibration events. Figure (15) shows a schematic of the power cycle treatment procedure. A small electronics fan was additionally programmatically controlled to speed the cooling phase of the power cycle.

\begin{figure}
	\centering
	\includegraphics[scale=1]{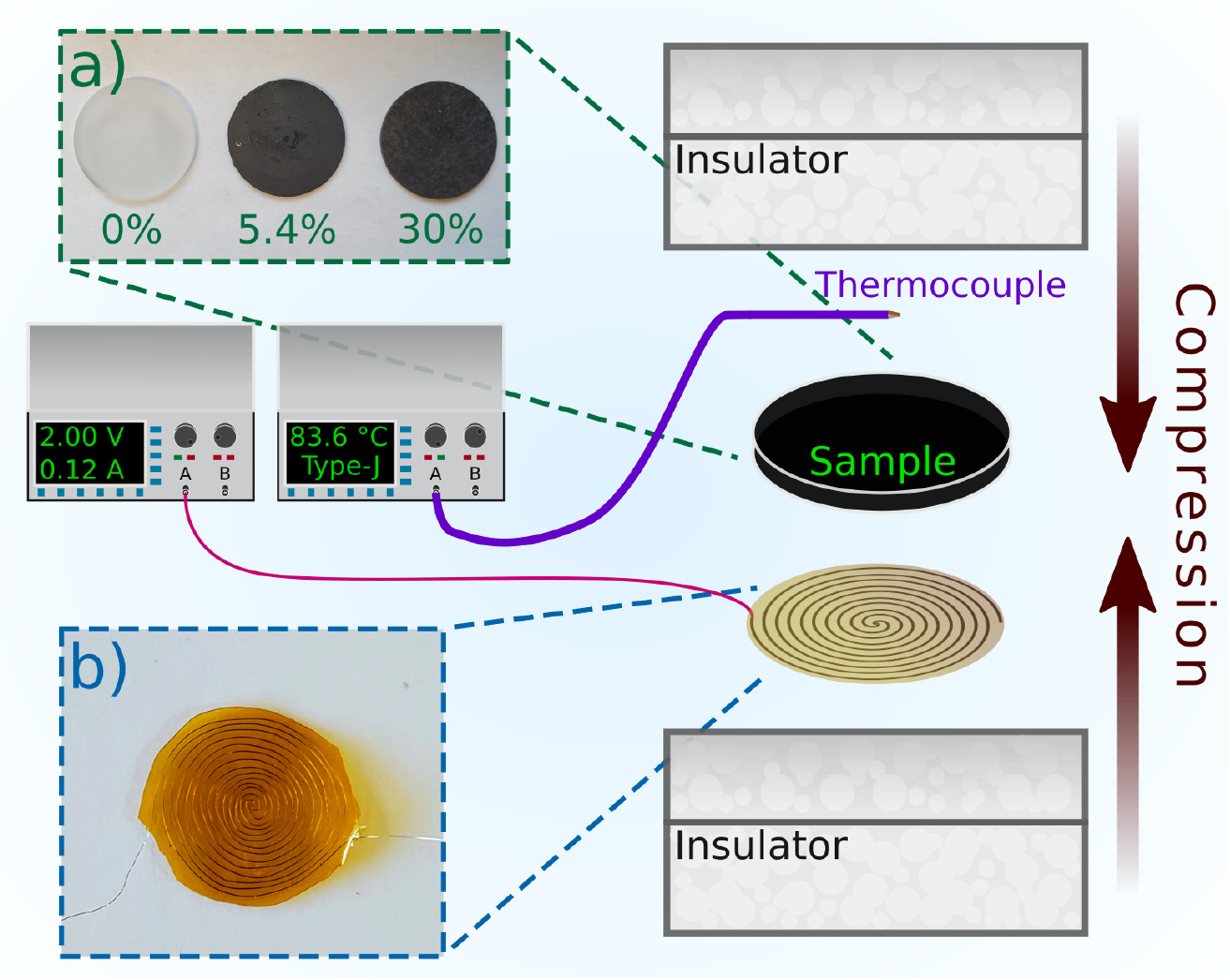}
	\caption{Schematic of the power cycling procedure. A power supply passes current through a custom wire coil heater and a thermocouple on back side of the sample measures the corresponding equilibrium temperature for that power level. The insulators, thermocouple, sample, and heating coil were all fixed in position with light compression. a) shows an image of the samples pre-treatment and b) shows an image of an example heating coil. Adapted with permission from ref \protect{\cite{Lewis2020}}. Published under the CC license by MDPI.}
\end{figure}

At specified power cycle counts, samples were removed from the power cycling apparatus and were experimented with LFA to directly measure their thermal diffusivity. The Figures (16a), (16b), shows the thermal diffusivities and conductivities for pure epoxy, while (16c) and (16d) shows that of 5.4 $vol. \%$, and (16e) and (16f) shows that of 30 $vol. \%$ samples. For all samples and at all power cycle counts, the thermal diffusivity reduced with increasing temperature. The initial RT diffusivities were 0.17, 1.25, and 4.6 mm$^{2}$/s, in order of increasing load level. After each sample's cycling treatments, their RT thermal diffusivities reached 0.17, 1.57, and 5.40 mm$^{2}$/s, in the same order, corresponding to a cycled percent enhancement of 0\%, 25.6\%, and 17.4\%. Interestingly, a clear increase in thermal diffusivity can be seen in loaded samples over the course of cycling. Though the pure epoxy sample does show modest improvement over the course of its cycling, it can only be seen at elevated temperatures, whereas the loaded samples show a more marked improvement at lower temperatures. 

\begin{figure}
	\centering
	\includegraphics[scale=0.8]{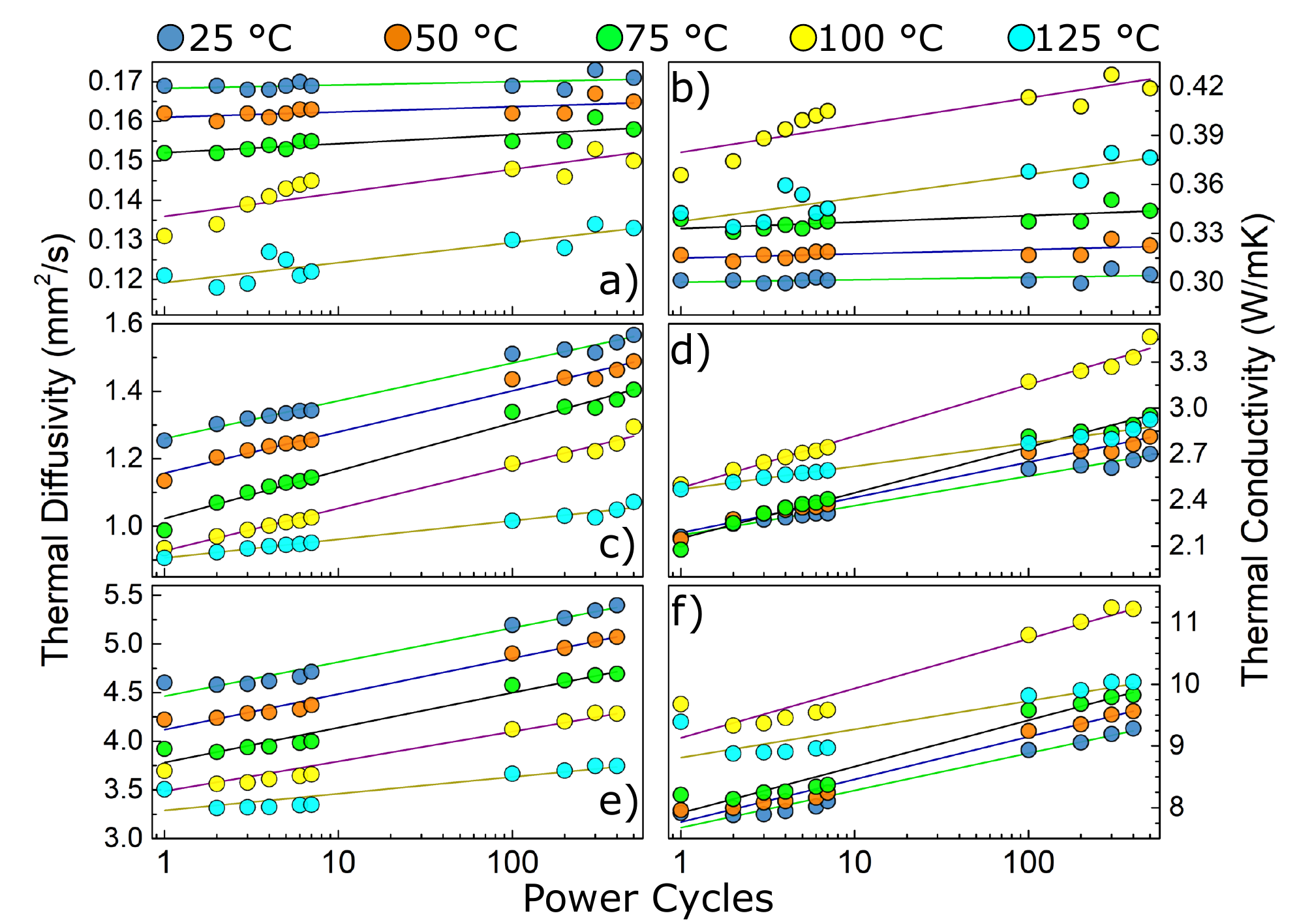}
	\caption{Panels on the left show thermal diffusivity and panels on the right show TC. a) Pure epoxy samples b) 5.4 $vol. \%$ samples c) 30 $vol. \%$ samples. Adapted with permission from ref \protect{\cite{Lewis2020}}. Published under the CC license by MDPI.}
\end{figure}

Using the definition of TC to be $K = \alpha \rho C_{p}$, LFA experiments for $\alpha$, Archimedes' Principle experiments for $\rho$, and the rule of mixtures for $C_{p}$ the TC of the composites were determined and is presented in Figures. After power cycling the 30 $vol. \%$ sample achieved a TC of 9.3 W/mK at RT, placing the sample among the highest reported for graphene-enhanced TIMs at this loading level \cite{Kargar2018, Kargar2019}. The pure epoxy sample did experience TC enhancement only past 100 $^{\circ}$C, with a modest enhancement of 7.7\% at that temperature, from 0.39 to $\approx$0.42 W/mK. However, the 5.4 $vol. \%$ and 30 $vol. \%$ samples each improved substantially over the course of cycling, constituting an improvement of 24.9\% and 17.3\%, respectively. The tendency for the TC of each composite to increase with temperature is primarily dictated by the composites' heat capacity behavior as temperature is varied. 

In each sample and at all cycle counts, the TC at 125 $^{\circ}$C is lower than at 100 $^{\circ}$C. This is attributed to the fact that the glass transition temperature of this material is at around 100 $^{\circ}$C and thermal properties are known to degrade in polymers beyond this temperature \cite{Kreahling1969, Park2015, DosSantos2013}. It has been reported previously that polymer glass transitions can be elevated with volumetric substitution of inert materials, such as graphene \cite{Yasmin2004, Ramanathan2008}. The appearance of a reduction in performance at 125 $^{\circ}$C indicates that any elevation of glass transition point must be less than 25 $^{\circ}$C in total. An increase of 30 $^{\circ}$C was seen previously in a PMMA polymer matrix with the inclusion of functionalized graphene. This suggests that graphene does not greatly inhibit the epoxy's cross-linking. 

No sample's performance decreased over the entire course of power cycling treatments. At low temperature, the pure epoxy sample's TC performance remained largely unchanged. Interestingly, the samples loaded with graphene, 5.4 $vol. \%$ and 30 $vol. \%$, showed a consistent increase in performance over the course of treatment. Due to the elimination of junction alterations as a factor to influence the TIM performance, the obtained results effectively present the intrinsic lifespan behavior of graphene TIMs in what should be a more reproducible experiment due to their being less potential variables. Should this study have been conducted in a junction it is very possible that the performance over the course of treatment would have decreased, as happened previously in a lifespan study of silver-filled epoxies \cite{Bjorneklett1993}.

Previously, accelerated lifespan study of a pure epoxy TIM showed modest thermal resistance reductions of 8\% indefinitely suggested to be caused by increasing the level of epoxy cross-linking \cite{Khuu2009}. Our study on pure epoxy is mostly in agreement with the previous results with only modest increases in TC -- which would accompany a reduction in thermal resistance -- and essentially no difference at temperatures below 100 $^{\circ}$C. When graphene is added to the epoxy, however, a clear increase in TC at all temperatures is observed over the course of power cycling, amounting to a percent enhancement of 24.9\% in 5.4 $vol. \%$ and 17.3\% in 30 $vol. \%$. Clearly from these results, graphene must play an essential role in the intrinsic TIM performance over the course of accelerated aging. 

It was reasoned that the increased cross-linking mechanic for enhanced performance proposed earlier could explain the large increase in graphene-epoxy TIMs but only modest increase in pure epoxy TIMs. If the epoxy matrix is increasing its level of cross-linking then it is swelling and simultaneously getting more and more rigid, leading to tighter mechanical coupling between graphene and epoxy matrix \cite{Kreahling1969}. This would lead to a lower Kapitza resistance between the two materials. As polymers are elevated in temperature the cross-linking rate can increase and once that reaction has taken place, it is irreversible with respect to temperature. This can explain why the over the course of power cycling the performance improves and why the improvement occurs even when tested at RT.

\section{Outlook}

TIMs play an important and increasing role in the behavior of high power electronic circuits and VLSI chips. Device miniaturization and densification is driving the unceasing demand for ever-improving TIM performance. Cost of production, ease of application, safety around exposed circuit elements, and lifespan reliability all contribute to the widescale adoption of polymer-based TIMs in industry. In order to improve the performance of polymer matrix TIMs, microscopic fillers of very high TC are dispersed within. Due to graphene's peerless intrinsic TC, it's advantageous quasi-2D geometry that traverses large distance in the polymer, and its broad face that strongly thermally couples the graphene to the matrix. Graphene's already impressive degree of coupling to the polymer matrix in which it is dispersed as well as its dispersibility can be further enhanced by functionalizing the graphene with numerous other materials. Tremendous potential for extraordinary TC enhancement exists for effective and facile alignment of graphene fillers. 

Increasing the load level of graphene in TIMs results in a super-linear TC enhancement past a point known as the thermal percolation threshold. The highest performance graphene TIMs of the future will very likely be past this load level, though special consideration will still need to be made towards the composite's viscosity and workability. In the percolative regime, graphene's cross-plane thermal conductivity plays a large role as a weakest link in overall heat flow.

Many applications of TIMs are very sensitive to its EC, whether the TIM directly encapsulates or is at risk to spilling onto active circuits due to junction pump-out as the device -- typically VLSI chip -- alternates between high and low power states. Because graphene has a high EC as well as TC, special care for electrically sensitive applications must be taken to either not surpass the electrical percolation threshold or to use a clever hybrid filler strategy to disrupt otherwise formed electrical networks within the composite. Dissimilarly shaped electrically insulating fillers have been shown to effectively result in a drastic reduction of majority graphene-filled TIMs by directly and contraposition studies.

TIMs by their nature must operate over a realistic lifespan. Any promising TIM developments require a full lifespan analysis before its industrial efficacy can be fully assessed. Unfortunately, there is little consistency among studies that are concerned with lifespan performance. It is recommended by the authors that a more simplified approach to TIM accelerated aging be taken in order to hopefully gain more consistency in research by reducing the number of testing parameters. 

\newpage
\footnotesize
\begin{longtable}{@{}p{0.17\textwidth}p{0.38\textwidth}p{0.12\textwidth}p{0.21\textwidth}p{0.13\textwidth}}
	
	\caption{TIM Thermal Conductivity Table. } \\
	
	\hline
	
	Base Polymer&Filler& Cross-plane TC (W/mK)&Measurement Method&Refs. \\
	
	\hline
	
	& \begin{flushleft} \textbf{Misc. Fillers} \end{flushleft} &  &  &  \\

	PDMS &  None &   0.2 &  ASTM D5470 &   \cite{Zhao2016} \\
	
	Polyolefin &  None &   0.3 & LFA &   \cite{Cui2015a} \\
	
	Epoxy & None & 0.2-0.22 & LFA &   \cite{Yu2007, Debelak2007} \\
	
	Olefin Oil &  None &0.145 & THW &   \cite{Choi2001} \\
	
	Mineral Oil &  None & 0.27-0.3 & ASTM D5470 &  \cite{Mahadevan2019, Naghibi2020} \\
	
	Epoxy &  None &  0.17-0.22 & LFA &  \cite{Kargar2018, Olowojoba2016, Yan2014} \\
	
	Silver Epoxy &  None &  1.67 & TPS &  \cite{Goyal2012} \\
	
	Paraffin &       None &       0.25 & TPS & \cite{Goli2014} \\
	
	Aerogel &       None &       0.18 & LFA & \cite{Zhong2013} \\
	
	Lauric Acid &       None &      0.215 & THW & \cite{Harish2015} \\
	
	Polyamide &       None &      0.196 & LFA & \cite{Ding2014} \\
	
	1-tetradecanol & None & 0.32 & TPS & \cite{Zeng2009} \\
	
	Commercial TIM & Undisclosed & 0.52-5.8 & ASTM D5470, LFA &\cite{Shahil2012, Kargar2014, Mahadevan2019} \\
	
	Commercial TIM & added h-BN 2 wt.\%/6 wt.\% &  0.56/.64 & ASTM D5470 & \cite{Kargar2014} \\
	
	Epoxy & AlN 60/74 vol.\% &   3.8/8.2 &  ASTM D5470 similar &   \cite{Ohashi2005} \\
	
	Epoxy & h-BN 43.6 vol.\% &  3.46 & LFA &  \cite{Lewis2019} \\
	
	Epoxy & h-BN 2.9 vol.\%/45 vol.\% &   0.32/5.5 & TPS, LFA &  \cite{Kargar2018} \\
	
	Epoxy & h-BN 15 vol.\% (CPA) &   6.1 &   TPS &  \cite{Han2019} \\

	Epoxy & h-BN 44 vol.\% & 9.0 & LFA &  \cite{Yu2017} \\
	
	Epoxy & h-BN 34 vol.\% &   4.4 & LFA &  \cite{Hu2017} \\
	
	Epoxy & h-BN 30 wt.\% &   0.6 & LFA &  \cite{Lin2014} \\
	
	Epoxy & h-BN 40 vol.\% (CPA) &   5.5 & LFA &  \cite{Kim2016c} \\
	
	Epoxy & h-BN 20 vol.\% &   1.2 & LFA &  \cite{Yuan2016} \\
	
	Epoxy & h-BN 50 vol.\% (Functionalized)&  9.81 & LFA &  \cite{Yu2015c} \\
	
	Epoxy & AlN 50 vol.\% &        1.21 &   TPS & \cite{Teng2012} \\

	Epoxy & Silica 50 vol. \% & 0.58 & ASTM E1530 & \cite{Wong1999} \\
	 
	Epoxy & SiC 72 wt.\% (Functionalized) &        5.75 & LFA & \cite{Zhou2013} \\
	
	Polyimide & h-BN 7 wt.\% &3 & LFA &  \cite{Wang2018} \\
	
	Polyimide & h-BN 60 wt.\% & 7.0 & TPS &  \cite{Sato2010} \\
	
	Polyimide & h-BN 60 wt.\% &  5.4 & TWA &  \cite{Tanimoto2013} \\
	
	Polyimide & h-BN 30 wt.\% &    0.72 & LFA & \cite{Song2019} \\
	
	PBT & h-BN 70 vol.\% (Functionalized) &    11 & LFA &  \cite{Morishita2016} \\
	
	PMMA & h-BN 80 wt.\% (Functionalized) &  10.2 & LFA &  \cite{Morishita2016} \\
	
	PCL & h-BN 20 wt.\% &  1.96 & LFA &  \cite{Lee2016} \\
	
	PVA & h-BN 30 wt.\% &  4.41 & LFA &  \cite{Xie2013} \\
	
	PVA & h-BN 10 wt.\% (Functionalized) &   5.4 & LFA &  \cite{Shen2015} \\	
		
	1-tetradecanol & Ag nanowires 11.8 vol. \% & 1.46 & TPS & \cite{Zeng2009} \\
	
	Silicone Oil & ZnO nanoparticles 18.7 vol. \% & 0.44 & TPS & \cite{Du2017} \\
	
	Silicone Oil & Zno Columns 18.7 vol. \% & 0.55 & TPS &  \cite{Du2017}\\
	
	Silicone Oil & ZnO Czech hedgehog structure 18.7 vol. \% & 0.83 & TPS & \cite{Du2017} \\
	 
	Resin & SiC 25 wt.\% &  1.28 & Unique Method &   \cite{Wang2006} \\

	& \begin{flushleft} \textbf{Non-Graphene Carbon Fillers} \end{flushleft} &  &  &  \\

	Epoxy & Small Graphite 4 wt.\%/13 wt.\%/20 wt.\% & 0.22/0.65/4.3 & LFA &   \cite{Debelak2007} \\
	
	Epoxy & Large Graphite 4 wt.\%/13 wt.\%/20 wt.\% & 0.87/2.95/4.3 & LFA &   \cite{Debelak2007} \\
	
	Epoxy & CF 20 wt.\% (non-heated/heated) &    0.35/3.75 & LFA &   \cite{Debelak2007} \\
	
	Epoxy & Graphite 10 wt.\% &   0.5 & LFA &  \cite{Song2013} \\
	 
	Epoxy & MWCNT 20 wt.\% &        0.4 & LFA & \cite{Yu2013} \\
	
	Epoxy & Graphite nanoplatelet (non/functionalized) 10 wt.\% & 0.65/1.75 & LFA & \cite{Zhou2013} \\
	
	Epoxy &Graphite 5.4 vol.\% (thicknesses 60 nm/30 nm/4 nm)&1.1/1.35/1.43&ASTM C518 &\cite{Yu2007}\\
	 
	Epoxy & Graphite Nanoplatelet 14 wt.\% &    0.73 & ASTM D5470 &   \cite{Tian2013} \\
	
	Silicone Oil & Graphite Nanoplatelet 14 wt.\% &   0.5 & ASTM D5470 &   \cite{Tian2013} \\
	
	Hatcol 2372 & Graphite Nanoplatelet 14 wt.\% &   0.48 & ASTM D5470 &   \cite{Tian2013} \\
	 
	Epoxy & SWCNT 1 wt. \% &        0.49 & ASTM D5470 similar & \cite{Biercuk2002} \\
	
	Epoxy & Graphite 44.3 wt. \% & 1.7 & TPS & \cite{Fu2014}\\
		
	Oil & MWCNT 1 vol.\% &    0.36 & THW &   \cite{Choi2001} \\
	
	CPE & SWCNT 50 wt.\% &   1.6 & TDTR &  \cite{Mai2016} \\
	
	Silver Epoxy & CB 5 vol.\% & 2 & TPS &  \cite{Goyal2012} \\

	& \begin{flushleft} \textbf{Graphene Fillers} \end{flushleft} &  &  &  \\

	Epoxy & GnP 20 wt.\% &        1.5 & LFA & \cite{Yu2013} \\
	
	Epoxy & Graphene 10 vol.\% &   5.1 & LFA &   \cite{Shahil2012} \\
	
	Epoxy & Graphene  11.4 vol.\%/43.6 vol.\% &   1.9/8.0 & LFA &  \cite{Lewis2019} \\
	
	Epoxy & Graphene 2.7 vol.\%/44.6 vol.\% &  0.49/11.4 & LFA &  \cite{Kargar2018} \\
	
	Epoxy & Graphene 55 wt.\% (Thicknesses 3 nm/ 12 nm) & 3.3/8 & LFA &  \cite{Kargar2019} \\
	
	Epoxy & Graphene 1 wt.\% (RA/CPA) &   0.2/0.35 & LFA &  \cite{Renteria2015} \\
	
	Epoxy & GnP 2 wt.\% (Functionalized) &   0.52 & LFA &  \cite{Chatterjee2012} \\
	
	Epoxy & Graphene 10 wt.\% (Functionalized) &   1.53 & LFA & \cite{Song2013} \\
	
	Epoxy & rGO 2 wt.\% &       0.24 & LFA & \cite{Olowojoba2016} \\
	
	Epoxy & Graphene 1 wt.\% (RA/CPA/IPA) &       0.4/0.57/0.25 & LFA & \cite{Yan2014} \\
	
	Epoxy & Graphene 0.92 vol.\% (CPA) &       2.13 & LFA & \cite{Lian2016} \\
	
	Epoxy & Graphene 10 wt.\% &       0.67 & LFA & \cite{Prolongo2014} \\
	
	Epoxy & Graphene 30 wt.\% &        4.9 & LFA & \cite{Tang2015} \\
	 
	Epoxy & Graphene 10 vol. \% & 3.35 & LFA & \cite{Dmitriev2018} \\
	
	Epoxy & GnP 8 wt.\%  &       1.18 & LFA & \cite{Wang2015c} \\
	
	Epoxy & GnP 10 wt.\%  &       6.5 & LFA & \cite{Moriche2016} \\
	
	Epoxy & Graphene alone/with PMMA 1 wt.\% &        0.6/1.4 & ASTM D5470 similar & \cite{Eksik2016} \\
	
	Epoxy & Graphene 5/10 vol.\% &        2.8/3.9 & ASTM D5470 similar & \cite{Park2015} \\
	 
	Epoxy & GnP 25 vol. \% & 6.75 & ASTM C518 & \cite{Yu2007} \\
	 
	Epoxy & Graphene 24 vol.\% &  12.4 &   DSC &  \cite{Shtein2015a} \\
	 
	Epoxy & Graphene 10.1 wt.\% & 4.0 &   TPS &  \cite{Fu2014} \\
	
	Polyamide & rGO wt.\% &      0.416 & LFA & \cite{Ding2014} \\
	
	Polyamide & rGO 5 wt.\% (Functionalized)&  0.41 &  &  \cite{Song2015} \\
	
	Polyamide & rGO 8 wt.\% (non/Functionalized) &        3.34/5.1 &        TPS & \cite{Cho2016} \\
	
	Polyurethane & rGO 1.04 wt.\% &        0.8 & LFA & \cite{Li2017a} \\
	
	Polyimide & Graphene 12 wt.\% &  0.41 & LFA &  \cite{Gong2016} \\
	
	Cellulose & rGO 30 wt.\% (IPA) & 0.07 & LFA &  \cite{Song2016} \\
	
	Mineral Oil & Graphene 10 wt.\%/20 wt.\%/40 wt.\% &  3.1/4.8/6.7 & ASTM D5470 &  \cite{Mahadevan2019} \\
	
	Mineral Oil & Graphene 27\% vol \% &   7.1 & ASTM D5470 &  \cite{Naghibi2020} \\
	
	Silver Epoxy & Graphene 1 vol.\%/5 vol.\% &   4.0/9.9 & TPS &  \cite{Goyal2012} \\
	
	Paraffin & Graphene 0.5 wt.\%/1 wt. \%/20 wt.\% &      10/15/45 & TPS & \cite{Goli2014} \\
	
	Commercial TIM & Added Graphene 2 wt.\%/4 wt.\%/6 wt.\% &0.7/0.75/0.8 &ASTM D5470&\cite{Kargar2014} \\
	
	Commercial TIM & Added Graphene 2 vol.\% &         14 & LFA & \cite{Shahil2012} \\
	
	Polystyrene & Graphene 20 wt.\% &   0.48 & LFA &  \cite{Cui2015b} \\
	
	Aerogel & rGO 20 vol.\% &       2.64 & LFA & \cite{Zhong2013} \\
	
	PDMS & Graphene  0.5 wt.\% (Scaffolded)& 0.4 &  ASTM D5470 & \cite{Zhao2016} \\
	
	Polyolefin & Graphene 10 wt.\% & 5.6 & LFA &   \cite{Cui2015a} \\
	
	Eicosane & Graphene 10 wt.\% &        2.0 & TPS & \cite{Fang2013} \\
	
	Lauric Acid & GnP 1 vol.\% &      0.49 & THW & \cite{Harish2015} \\
		
	Methyl Vinyl Silicone & rGO 1.5 wt.\% &        2.7 & LFA & \cite{Zhang2016b} \\
	
	PVDF & rGO 0.25 wt. \% & 2.35 & LFA & \cite{Kumar2019} \\ 

	& \begin{flushleft} \textbf{Hybrid Fillers} \end{flushleft} &  &  &  \\
	
	Epoxy & Graphene 21.8 vol.\%, h-BN 21.8 vol.\% &   6.5 & LFA &  \cite{Lewis2019} \\
	
	Epoxy & GnP 40 wt.\%, Cu-NP 35 wt.\% & 13.5 & LFA &  \cite{Barani2020} \\
	 
	Epoxy & MWCNT grown on GnP 20 wt.\% &        2.4 & LFA & \cite{Yu2013} \\
	 
	Epoxy & AlN nanowires 30 vol.\%, AlN spheres 30 vol.\% &        5.23 & LFA & \cite{Dang2017} \\
	 
	Epoxy & BN nanowires 12.8 vol.\%, BN spheres 30 vol.\% &        3.6 & LFA & \cite{Kim2014}\\
	 
	Epoxy & Al$_{2}$O$_{3}$-attached GnP 12 wt.\% &        1.49 & LFA & \cite{Sun2016b} \\
	 
	Epoxy & Ag-attached h-BN 25.1 vol.\% &        3.1 & LFA & \cite{Wang2016c} \\
	 
	Epoxy & Graphene oxide 49.6 wt.\%, MWCNT 0.4 wt.\% &        4.4 & LFA & \cite{Im2012} \\
	 
	Epoxy & h-BN, SiC 40 vol.\% total (CPA) &        5.77 & LFA & \cite{Kim2016d} \\
	 
	Epoxy & h-BN, rGO 13.2 wt.\% total (CPA) &        5.1 & LFA & \cite{Yao2018} \\
	 
	Epoxy & MWCNT 5 wt.\%, SiC 55 wt.\% &        6.8 &     LFA    & \cite{Zhou2010} \\
	 
	Epoxy & AlN 40.9 wt.\%, Al$_{2}$O$_{3}$ 17.5 wt.\% &        3.4 &     LFA    & \cite{Choi2013} \\
	 
	Epoxy & rGO 20 wt.\%, Graphene 10 wt. \% (Scaffolded) & 6.7 & LFA & \cite{Tang2018} \\
	 
	Epoxy & AlN 25 vol.\%, MWCNT 1 vol.\% &        1.21 &   TPS & \cite{Teng2012} \\
	 
	Epoxy & Graphene oxide 6 wt.\%, AlN 50 wt.\% &        2.77 &     TPS    & \cite{Yuan2016b} \\
	 
	Epoxy & MWCNT 4 wt.\%, AlN 25 wt.\% &          1 &  TPS  & \cite{Ma2012} \\ 
	 
	Epoxy & MWCNT 15 wt.\%, Cu 40 wt.\% &        0.6 &  TPS  & \cite{Zhang2014} \\
	 
	Epoxy & Graphene 0.9 wt.\%, MWCNT 0.1 wt.\% &        0.3 & TPS & \cite{Yang2011} \\ 
	
	Epoxy & Silica-coated AlN 50 vol. \% & 1.96 & ASTM E1530 & \cite{Wong1999} \\
	
	Epoxy & Graphene 16 vol.\%, h-BN 1 vol.\% &   4.72 & DSC &  \cite{Shtein2015a} \\
	
	Epoxy & Ag nanowires 4 vol.\%, Al$_{2}$O$_{3}$ 15 wt.\% &          1.08 & TPS similar & \cite{Chen2016} \\ 

	Epoxy & Graphene 1.5 wt\%, MgO 30 wt.\% &        0.51 & ASTM D5470 similar & \cite{Liu2019} \\ 

	Epoxy & MgO-coated Graphene 7 wt.\% &        0.4 & ASTM C518 & \cite{Du2015} \\ 

	Epoxy & Al$_{2}$O$_{3}$ 30 wt.\%, rGO 0.3 wt.\% &        0.33 &        ASTM E1461 & \cite{Zeng2015} \\

	Epoxy & Graphene oxide-encapsulated h-BN 40 wt.\% &        2.2 & ASTM D5470 & \cite{Huang2016b} \\
	 
	Polyimide & h-BN ($\mu$m scale) 21 wt. \% h-BN (nm scale) 9 wt.\% &        1.2 &   TPS& \cite{Li2010} \\
	 
	Polyimide  & BN-coated Cu Nanoparticles, Nanowires 10 wt.\% total &        4.3 & TPS & \cite{Zhou2018} \\
	 
	Polyimide & BN 50 wt.\%, Graphene 1 wt.\% &  2.11 &  ASTM D5470 &  \cite{Tsai2014} \\
	
	Polyamide & Graphene 20 wt.\%, h-BN 1.5 wt.\% &        1.76 & LFA & \cite{Cui2015b} \\ 
	
	Polyamide & Graphene oxide 6.8 wt.\%, h-BN 1.6 wt.\% &        0.9 & LFA & \cite{Shao2016} \\ 
	
	Polycarbonate & GnP 18 wt.\%, MWCNT 2 wt.\% &        1.39 &   TPS & \cite{Yu2016} \\
	
	PDMS & Graphene (Scaffolded), CB 2 wt.\%/8 wt.\% &  0.41//0.7 &  ASTM D5470 &   \cite{Zhao2016} \\

	PPS & h-BN ($\mu$m scale) 40 wt.\%, h-BN (nm scale) 20 wt.\% &  2.64 & TPS & \cite{Gu2017} \\
	
	PPS & h-BN 50 wt.\%, MWCNT 1 wt.\% &        1.74 & TPS similar & \cite{Pak2012} \\

	PVA & Graphene, MWCNT each Ag-attached 20 vol.\% total &  12.3 & LFA &  \cite{Zhou2018b} \\
	
	Polystyrene & GnP 20 wt.\%, h-BN 1.5 wt.\% &        0.66 & LFA &  \cite{Cui2015b}   \\
	
	PVDF & GnP 5 wt.\%, Nickel 8 wt.\% &        0.66 & LFA & \cite{Zhao2018} \\

	Cyanate Ester & Graphene 5 wt.\%, iron-nickel alloy 15 wt.\% &        4.1 &        TPS & \cite{Ren2018} \\

	Polylactic acid & Alumina 70 wt.\%, graphene 1 wt.\% & 2.4 & TPS & \cite{Jiang2020}\\

     \hline
\end{longtable}

\begin{longtable}{@{}p{0.2\textwidth}p{0.5\textwidth}}
	\caption{Table Acronym Legend} \\
	\hline
	Acronym & Meaning \\
	\hline
	
	PDMS& Poly(dimethylsiloxane)\\
	LFA & Laser Flash Analysis \\
	TPS & Transient Plane Source \\
	``X similar" & Shares similarities to X\\
	RA  & Randomly aligned filler (Studies without any alignment classification are randomly oriented)\\
	CPA & Cross-plane filler preferential alignment\\
	IPA & In-plane filler preferential alignment\\
	TWA & Temperature Wave Analysis\\
	PBT & Polybutylene terephthalate\\
	PCL & Poly(caprolactone)\\
	PVA & Poly(vinyl alcohol)\\
	PMMA& poly(methyl methacrylate)\\
	CF  & Carbon fiber\\
	MWCNT&Multi-walled carbon nanotube\\
	SWCNT&Single-walled carbon nanotube\\
	THW & Transient Hot Wire\\
	CPE & Conjugated polyelectrolytes\\
	TDTR& Time-domain Thermoreflectance\\
	CB  & Carbon black\\
	GnP & Graphene nanoplatelet\\
	rGO & Reduced graphene oxide\\
	PVDF& Poly(vinylidene fluoride)\\
	PPS & Poly(phenylene sulfide)
\end{longtable}
\normalsize
\newpage
\bibliography{TimReviewBib}
\bibliographystyle{unsrt}
\end{document}